\documentclass[a4paper,fleqn]{cas-dc}

\usepackage[numbers]{natbib}

\usepackage{algorithm}
\usepackage{algorithmicx}
\usepackage[noend]{algpseudocode}

\usepackage{tablefootnote}

\def\tsc#1{\csdef{#1}{\textsc{\lowercase{#1}}\xspace}}
\tsc{WGM}
\tsc{QE}

\begin{document}
\let\WriteBookmarks\relax
\def\floatpagepagefraction{1}
\def\textpagefraction{.001}

\shorttitle{Turn Tree into Graph: Automatic Code Review via Simplified AST Driven Graph Convolutional Network}    

\shortauthors{Wu et~al.}  

\title [mode = title]{Turn Tree into Graph: Automatic Code Review via Simplified AST Driven Graph Convolutional Network}  



%


\author[1]{Bingting Wu}[type=editor, style=chinese,]
\ead{20204227028@stu.suda.edu.cn}

\affiliation[1]{organization={School of Computer Science and Technology, Soochow University},
            city={Suzhou},
            country={China}}

\affiliation[2]{organization={Department of Computer Science, Harbin Institute of Technology},
            city={Shenzhen},
            country={China}}

\author[2]{Bin Liang}[style=chinese]
\cormark[1]
\ead{bin.liang@stu.hit.edu.cn}

\author[1]{Xiaofang Zhang}[style=chinese]
\cormark[1]
\ead{xfzhang@suda.edu.cn}

\cortext[cor1]{Corresponding author.}



\begin{abstract}
Automatic code review (ACR), which can relieve the costs of manual inspection, is an indispensable and essential task in software engineering.
To deal with ACR, existing work is to serialize the abstract syntax tree (AST). However, making sense of the whole AST with sequence encoding approach is a daunting task, mostly due to some redundant nodes in AST hinder the transmission of node information. Not to mention that the serialized representation is inadequate to grasp the information of tree structure in AST.
In this paper, we first present a new large-scale Apache Automatic Code Review (AACR) dataset for ACR task since there is still no publicly available dataset in this task. The release of this dataset would push forward the research in this field.
Based on it, we propose a novel Simplified AST based Graph Convolutional Network (SimAST-GCN) to deal with ACR task.
Concretely, to improve the efficiency of node information dissemination, we first simplify the AST of code by deleting the redundant nodes that do not contain connection attributes, and thus deriving a Simplified AST.
Then, we construct a relation graph for each code based on the Simplified AST to properly embody the relations among code fragments of the tree structure into the graph.
Subsequently, in the light of the merit of graph structure, we explore a graph convolution networks architecture that follows an attention mechanism to leverage the crucial implications of code fragments to derive code representations. Finally, we exploit a simple but effective subtraction operation in the representations between the original and revised code, enabling the revised difference to be preferably learned for deciding the results of ACR.
Experimental results on the AACR dataset illustrate that our proposed model outperforms the state-of-the-art methods.
\end{abstract}




\maketitle





\section{Introduction}
Code review is the act of consciously and systematically convening programmers to check each other’s code for mistakes, and has been repeatedly shown to accelerate and streamline the process of software development. Hence, it also incurs considerable human resources \cite{sadowski2018modern}, making it impossible to expand code review on a large scale. Therefore, many researchers are committed to automatic code review (ACR). For ACR, we first provide the model with the original code and the revised code, and then the model provides us with suggestions on whether this modification is acceptable.

Traditional approaches are limited to deal with the main challenge of code review: understanding the code \cite{bacchelli2013expectations}. Therefore, researchers can only improve efficiency from other aspects of code review, such as recommending suitable reviewers \cite{thongtanunam2015should,zanjani2015automatically,xia2017hybrid} and using static analysis tools \cite{balachandran2013reducing,diaz2013static,mcgraw2008automated}. However, with the development of deep learning, we can understand the code by modeling the source code, thereby effectively solving the main challenges in automatic code review.

Recent work \cite{shi2019automatic,siow2020core} has shown that deep learning methods perform better in capturing the syntactical and semantic information of the source code, enabling suitable code review suggestions. Among them, Shi \cite{shi2019automatic} proposes a method called Deep Automatic Code reviEw (DACE), which uses long short-term memory (LSTM) \cite{greff2016lstm} and  convolutional neural network (CNN) \cite{Sun_2021_CVPR} to capture the semantic and syntactical information, respectively. Due to the characteristics of ACR, the model needs to compare the original code and the revised code. They also design a pairwise recursive autoencoder to compare the code.

In most previous research efforts, they divide the code according to the delimiter to ensure that the syntactical information of the code can be preserved. However, such delimiter-based models still limit in that splitting the code according to the delimiter does not effectively represent the structural information of the code. This is because of the differences between programming languages and natural languages. In natural language, it is usually sequential comprehension. But in a programming language, it needs to be understood under the logical order of the abstract syntax tree (AST). For example, the programmer may divide the code into lines because the code is too long, but this does not mean that the syntax of the code has changed. 

In the field of code representation, moreover, there are some AST-based methods. These methods mostly serialize the AST into a sequence of nodes. In the subsequent processing, various network models can be applied to improve the performance of code representation. Although these methods use some structural information in the AST, they do not make full use of the structural information at the model level.

We thus explore a novel solution to represent the code fragments: obtaining a more concise and efficient code graph representation by simplifying the AST and using graph convolution to handle the associated information between nodes. Based on the idea, we propose a Simplified AST based Graph Convolutional Network (SimAST-GCN) model to leverage the semantic dependencies of the code fragments. Here, the semantic and syntactical information from neighbors of each node are aggregated to derive the code graph embeddings, so as to extract the semantics for representing the code fragments well. To our knowledge, this is the first study to deploy the graph structure for leveraging the node connection information in the AST for the code review task. Further, there is no public dataset for the code review task. As such, to advance and facilitate research in the field of ACR, we present a new Apache Automatic Code Review (AACR)
dataset. The main contributions of our work can be summarized as follows:

\begin{itemize}
\item We provide a large-scale Apache Automatic Code Review dataset (AACR), since there is no public dataset available for ACR.
\item A Simplified AST based Graph Convolutional Network is proposed to extract syntactical and semantic information from source code fragments.
\item Experimental results on the AACR dataset show that the proposed model achieves significantly better results than state-of-the-art baseline methods.
\end{itemize}

The rest of this paper is organized as follows. Section 2 introduces the background. 
Section 3 describes our approach.
Section 4 provides our experimental design.
Section 5 presents the experimental results and analyzes them.
Section 6 presents several related works.
Finally, Section 7 concludes our work.

\begin{figure}[t]
    \centering
    \includegraphics[width=80mm]{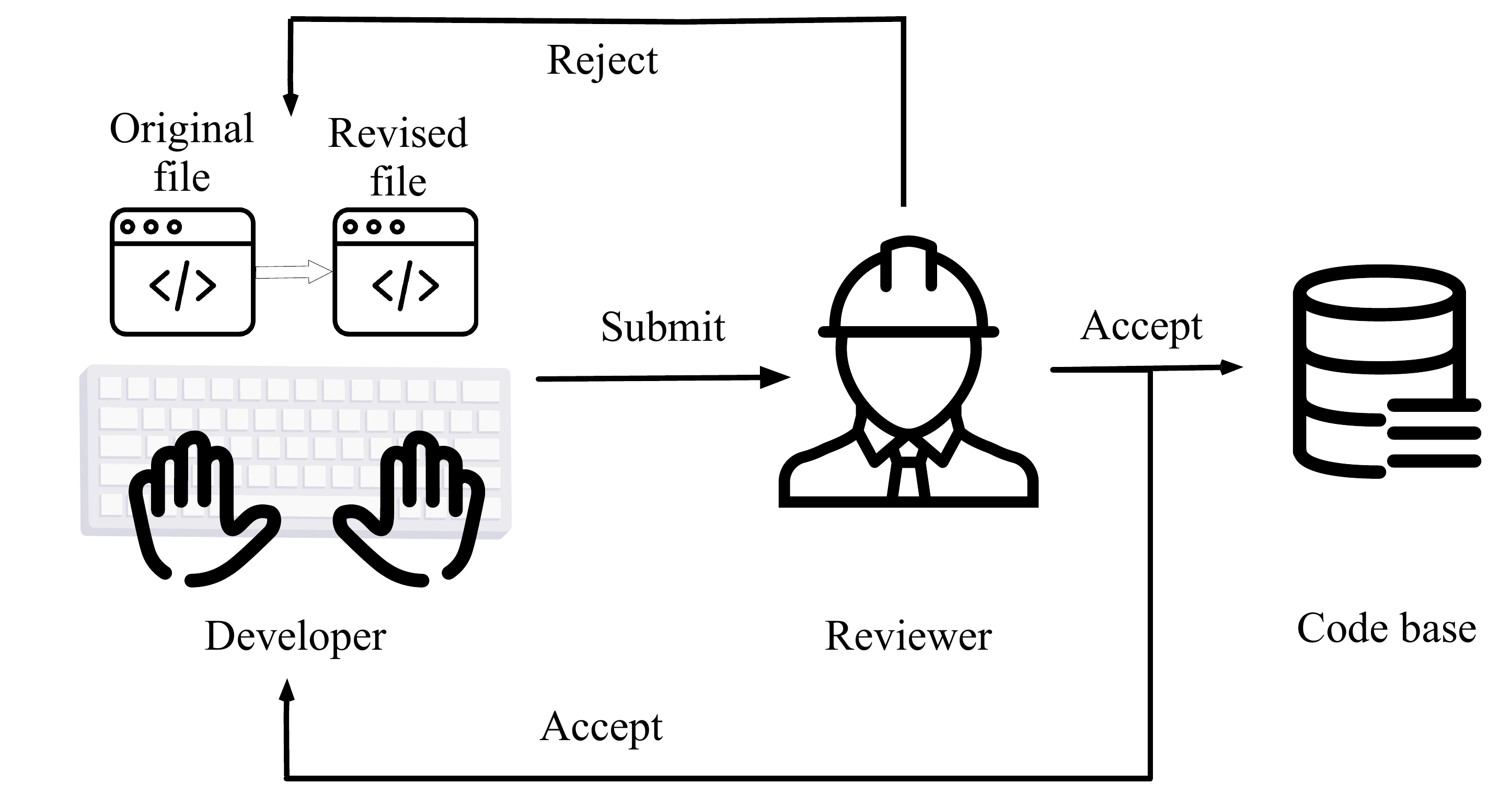}\\
    \caption{Traditional code review process.}
    \label{fig:cr}
\end{figure}
  
\section{Background}
\subsection{Code Review}
The general process of traditional code review is shown in Figure 1. When a developer completes and submits the code implementation for specific requirements, the system will arrange a suitable reviewer who needs to compare the differences between the original file and the revised file to verify whether the code meets the requirements. If there is no problem, the code will be added to the code base; otherwise, the reviewer will ask the developer to revise the code.

Traditional approaches use static analysis tools to assist with code review. For example, Checkstyle\footnote{https://checkstyle.sourceforge.io} covers coding style-related issues, PMD\footnote{https://pmd.github.io} checks class design issues and questionable coding practices, and FindBugs\footnote{http://findbugs.sourceforge.net} \cite{hovemeyer2004finding} detects potential bugs in the code. However, traditional static analysis tools cannot understand the code. They only judge whether there is a problem with the code based on predefined patterns.

To solve the ACR task with a deep learning method, the model first extracts as many features as possible from the original file and the revised file, and encode these features as vector representations. Then, the model uses different network structures to enhance these features to maximize the characteristics of the source code. Finally, it is necessary to design a suitable model or method to calculate the distance between the original file and the revised file, and generate code review recommendations based on the difference.

\begin{figure}[t]
    \centering
    \includegraphics[width=0.68\linewidth]{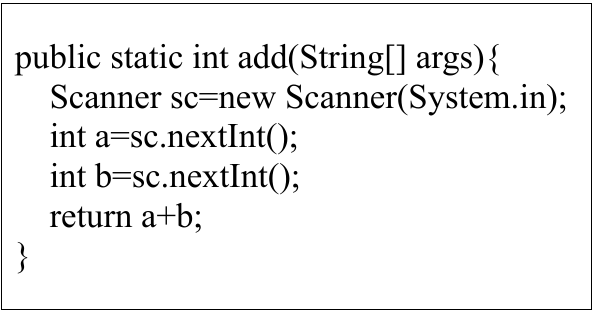}\\
    \caption{Example of source code.}
    \label{fig:code}
  \end{figure}

\begin{figure*}[t]
    \centering
    \includegraphics[width=1\linewidth]{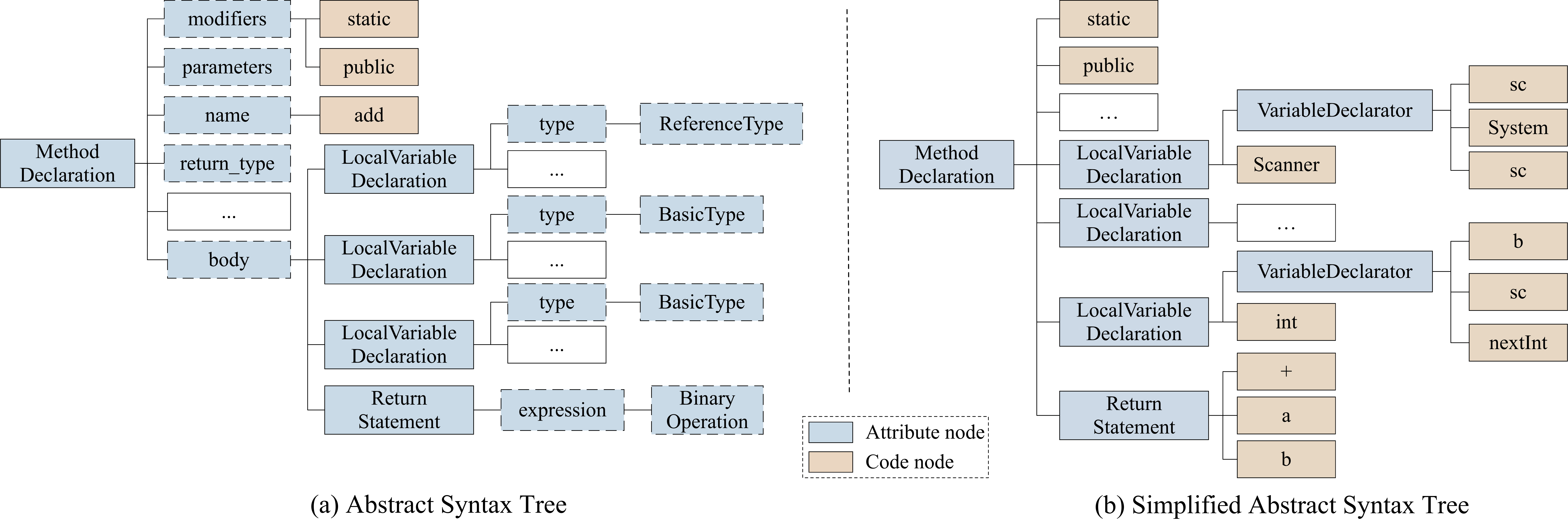}\\
    \caption{Example of preprocessing source code based on AST.  The nodes framed by dashed lines are later removed. The AST is extracted from the source code shown in Figure 2.}
    \label{fig:ast}
  \end{figure*}

\subsection{Abstract Syntax Tree}
An abstract syntax tree (AST) is a tree designed to represent the abstract syntactic structure of source code \cite{baxter1998clone}. For example, Figure 2 shows an example of source code, and Figure 3(a) shows the AST extracted from that source code. AST has been widely used by programming languages and software engineering tools. For example, it has a wide range of applications in source code representation \cite{8812062,mou2014tbcnn}, defect prediction \cite{shippey2019automatically}, and other fields. Each node of an AST corresponds to a construct or symbol in the source code. Firstly, unlike ordinary source code, AST is abstract and does not contain all the details, such as separators and punctuation. Secondly, AST contains richer structural and semantic information, which is very suitable for representing source code. 

Here, we do not directly use the original AST information. Since after the program is parsed into an AST, its node size will increase significantly, which hampers the performance of the graph network. Therefore, we simplify the AST to improve the performance of the entire model.

\subsection{Tree-based Neural Network}
Recently, many researchers have proposed Tree-Based Neural Networks (TBNNs) that use ASTs directly as model inputs. Given a tree, TBNNs typically use a recursive approach, aggregating information from leaf nodes upward in layers to obtain a vector representation of the source code. The most representative model of the code representation is AST-based Neural Network (ASTNN) \cite{8812062}.

In ASTNN, they parse the code fragment into the AST and use the preorder traversal algorithm to split the AST into a sequence of statement trees (ST-trees, that is, a tree composed of the statement node as the root and the AST node corresponding to the statement). Then, they design a Statement Encoder module to encode all ST-trees into vectors $e_1,...,e_t$. For each ST-tree $t$, let $n$ denotes a non-leaf node, and $C$ denotes the number of its children nodes. With the pre-trained embedding matrix $W_e \in \mathbb{R}^{|V| \times d}$ where $V$ is the vocabulary size and $d$ is the embedding dimension of nodes, the vector of node $n$ can be obtained by:
\begin{equation}
	v_n = W_e^\top x_n
\end{equation}
where $x_n$ is the one-hot representation of symbol $n$ and $v_n$ is the embedding. Next, the representation of node $n$ is computed by the following equation:
\begin{equation}
	h=\sigma(W_n^\top v_n+\sum_{i\in[1,C]} h_i + b_n)   
\end{equation}
where $W_n \in \mathbb{R}^{d \times k}$ is the weight matrix with encoding dimension $k$, $h_i$ is the hidden state for each children $i$, $b_n$ is a bias term, $\sigma$ is the activation function and $h$ is the updated hidden state. The final representation of the ST-tree is computed by:
\begin{equation}
	e_t = [max(h_{i1}),...,max(h_{ik})],i=1,...,N    
\end{equation}
where $N$ is the number of nodes in the ST-tree. Then, ASTNN uses Bidirectional Gated Recurrent Unit (Bi-GRU) \cite{tang2015document}, to model the characteristics of the statements. The hidden states of Bi-GRU are sampled into a single vector by pooling. Such a vector captures the characteristics of source code \cite{hindle2016naturalness}, \cite{ray2016naturalness} and can serve as a neural source code representation.

\subsection{Motivation}
As illustrated in the previous sections, the structural information of code, like AST, Control Flow Graph (CFG) \cite{fang2020functional}, Data Flow Graph (DFG) \cite{guo2020graphcodebert}, have been widely used in many tasks. We think that it's meaningful to use the structural information in ACR task.

For token-based methods and delimiter-based methods, the performance is under the expectation in many tasks. Their approaches are to divide the code into tokens or lines according to the space or delimiter. There is no doubt that these pre-processing operations abandon the structural information. However, in the code representation, the structural information is much more important than that in nature language process. Hence, it is a little late for us to use the AST information in ACR task.

In addition to introducing AST structure information into ACR, we also made some optimizations and improvements. In many code representation methods, we noticed that many researchers \cite{8812062} believe that the large number of AST nodes has a negative effect on the model.

In order to solve this problem, we manually checked all the node types generated by AST, and tried to filter the generated nodes with simple rules. Attempt to greatly reduce the number of automatically generated nodes without affecting the overall structure and semantic information of the AST, so as to obtain lightweight and effective structural information.

Similarly, we also found that in the field of code representation, there are still some shortcomings in the use of AST's tree structure information. Researchers usually serialize the tree structure into a sequence of nodes, which damages the overall information expression to some extent. Therefore, in this article, we also propose the use of the latest graph convolution operation and attention mechanism to better capture code information from the tree structure.

\begin{figure*}[t]
    \centering
    \includegraphics[width=1\linewidth]{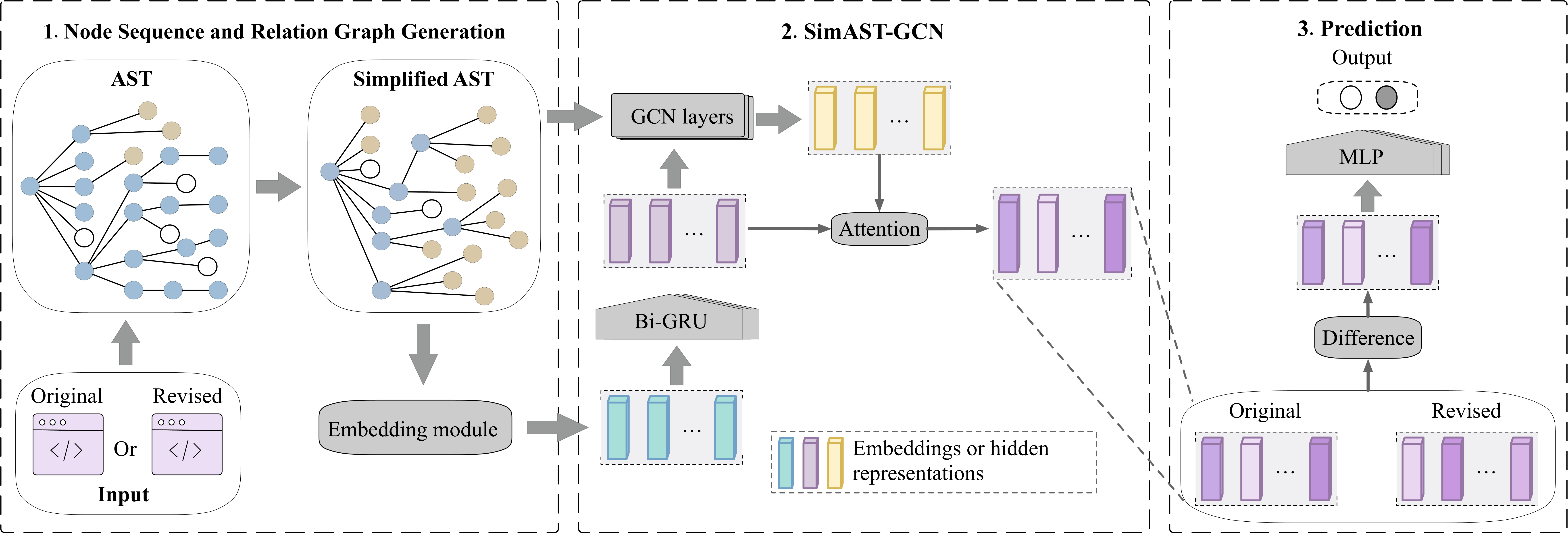}\\
    \caption{General framework of SimAST-GCN. The blue and khaki nodes in the AST correspond to the nodes in Figure 3. The white nodes are the undrawn parts.}
    \label{fig:model}
\end{figure*}

\section{Proposed Approach}
We introduce our approach (SimAST-GCN) in this section. As shown in Figure 4, the architecture of the proposed SimAST-GCN framework contains three main components. First, the Node Sequence and Relation Graph Generation module simplifies the AST, as illustrated in Figure 3, and generates the corresponding adjacency matrix and node sequence based on the Simplified AST. Second, the proposed SimAST-GCN obtains the word embeddings of the node sequence and uses a Bidirectional Gated Recurrent Unit (Bi-GRU) to model the naturalness of the statements. It then employs a Graph Convolution Network (GCN) to fuse the node relation graph and the hidden status. Finally, retrieval-based attention is employed to derive the code representation. Third, a prediction is made by calculating the differences in the code representations to predict the final result.

\subsection{Node Sequence and Relation Graph Generation}

\subsubsection{Simplifying AST}
First, we use the existing syntax analysis tools to parse the source code fragments into the corresponding ASTs. For each AST, we delete the redundant nodes and reconstruct node connections of the entire AST to ensure the integrity of the tree structure.

Given an AST $T$ and the AST attribute nodes $S$ (attribute nodes, the blue nodes in Figure 3). First, we filter attribute nodes and retain nodes with strong semantic information. We assume that if the attribute nodes contain a \emph{Declaration} or \emph{Statement}, then these attribute nodes contain connection information. For example, in Figure 3(a), \emph{MethodDeclaration} defines a method, all nodes under the method belong to this node, while the node \emph{modifiers} indicates that the node \emph{static} is a modifier, and reduces the strength of the connection between the node \emph{static} and \emph{MethodDeclaration}. Therefore, we remove these redundant nodes to reduce the number of node sequences and increase the strength of the connection between each node. 

If we simply remove the redundant nodes in the AST tree, it will split the entire AST. To maintain the integrity of the whole tree, we need to reconnect the split AST. For example, in Figure 3(a), the nodes framed by dashed lines are removed, and finally a Simplified AST is generated, as shown in Figure 3(b). The procedure for simplifying the AST is depicted in Algorithm 1. In general, if the node in the original AST is deleted, we connect the child node of the deleted node to its parent node.

\begin{algorithm}[tb]
\caption{Procedure for simplifying the AST}
\label{alg:algorithm}
\textbf{Input}: The root of AST, R; the source code fragment, C; the attribute node, S.\\
\textbf{Output}: The Simplified AST.
\begin{algorithmic}[1] 
\State Let $FS = [\text{ }]$.
\For{each $node \in S$} 
\If {$Declaration \in node$ or $ Statement \in node$}
\State $FS \leftarrow FS + node$
\EndIf
\EndFor

\Function{SimplifyAST}{$R, C, FS$}
    \State Let $children = [\text{ }]$.
    \For{each $ node \in R[child]$}
    \If {$ node \in C$ or $node \in FS$}
    \State $children \leftarrow children + node$
    \Else \State $children \leftarrow children + node[child]$
    \EndIf
    \State $SimplifyAST(node, C, FS)$
    \EndFor
    \State $R[child] \leftarrow children$
\EndFunction

\State \Call{SimplifyAST}{$R, C, FS$}
\State \Return $R$
\end{algorithmic}
\end{algorithm}

\subsubsection{Generate Node Sequence and Relation Graph over Simplified AST}
After obtaining the Simplified AST, we use the depth-first traversal algorithm to serialize the Simplified AST into Node Sequences. For example, if the size of the AST is $n$, then we derive a node sequence $w = [w_1,w_2,...,w_n]$.

Inspired by previous GCN-based approach \cite{zhang2019aspect}, we produce a node relation graph for each code fragment over the simplified AST:
\begin{equation} 
    A_{i,j} = \begin{cases}

        1& \text{if $i=j$ or $w_i,w_j$ are directly connected}\\
        
        0& \text{otherwise}
        
        \end{cases}
\end{equation}
Then, an adjacency matrix $\boldsymbol{A}\in \mathbb{R}^{n\times n}$ is derived via the simplified AST of the source code fragments.

\subsection{SimAST-GCN}

\subsubsection{Embedding Model}
In our proposed model SimAST-GCN, we use the gensim \cite{rehurek_lrec} library to train the embeddings of symbols to get the distributed representations of the words in the AST. Thus, we can get the embedding lookup table $\boldsymbol{V}\in \mathbb{R}^{m\times |N|}$ according to the word index, where $m$ is the embedding size (the dimension of each word) and $|N|$ is the number of all words after deduplication (vocabulary size). Then, given a node sequence with $n$ nodes, we can get the corresponding embedding matrix $\mathbf{x}=[\mathbf{x_1},\mathbf{x_2},...,\mathbf{x_n}]$, where $\mathbf{x_i} \in \mathbb{R}^m$ is the word embedding.

\subsubsection{Graph Convolutional Network}
In our model SimAST-GCN, our GCN module takes the node relation graph and the corresponding node representations as input. Each node representation in the $l$-th GCN layer is updated by aggregating the information from their neighborhoods, the calculation formula is:
\begin{equation}
    \mathbf{h}^l = \mathrm{LeakyReLU}(\mathbf{L} \mathbf{h}^{l-1} \mathbf{W}^l + \mathbf{b}^l)
\end{equation}
where $\mathbf{h}^{l-1}$ is the hidden representation generated from the preceding GCN layer. $\mathbf{L}$ is a normalized symmetric of a node relation adjacency matrix:
\begin{equation}
    \mathbf{L} = \mathbf{A} / (\mathbf{D} + 1)
\end{equation}
where $\mathbf{D} = \sum\nolimits_{j=1}^nA_{i,j}$ is the degree of $\mathbf{A}_i$. The original node representations for the GCN layers are the hidden representations generated by the Bi-GRU layers, which using the previous embedding matrix $\mathbf{x}$ as the $1-$st GCN layer input:
\begin{equation}
    \mathbf{H}^c = \{\mathbf{h}_1^c,\mathbf{h}_2^c,...,\mathbf{h}_n^c\} = \text{Bi-GRU}(\mathbf{x})
\end{equation}
Finally, we can capture the representations $\tilde{\mathbf{h}}$ of the GCN layers successfully. Susequently, we use the retrieval-based attention mechanism \cite{zhang2019aspect} to capture significant sentiment features from the context representations for the source code:
\begin{equation}
    \beta_t = \sum_{i=1}^n \mathbf{h}_t^{c\top} \tilde{\mathbf{h}}_i
\end{equation}
\begin{equation}
    \alpha_t = \cfrac{\mathrm{exp}(\beta_t)}{\sum\nolimits_{i=1}^{n} \mathrm{exp}(\beta_i)}
\end{equation}
where $\mathbf{h}_t^{c\top}$ is the transposition of the hidden status of the $t$-th node, and $\tilde{\mathbf{h}}_i$ is the graph hidden representation of the $i$-th node. Hence, the final representation of the source code fragment is formulated as follows:
\begin{equation}
    \mathbf{r} = \sum_{i=1}^n \alpha_i \mathbf{h}_i^c
\end{equation}

\subsection{Prediction}
The above content is the operation for one code fragment. The ACR process needs to compare the two source code fragments (the original file $\mathbf{s}^O$ and the revised file $\mathbf{s}^R$) and give a judgment---that is, whether it passes the code review. Therefore, after we obtain the corresponding representations $\mathbf{r}^O$ and $\mathbf{r}^R$, we need to calculate the distance between them:
\begin{equation}
    \mathbf{r} = \mathbf{r}^O-\mathbf{r}^R
\end{equation}
\begin{equation}
    \mathbf{y} = \mathrm{softmax}(\mathbf{W}  \mathbf{r} + \mathbf{b}) 
\end{equation}
where $\mathrm{softmax}(\cdot)$ is the softmax function.

The target to train the classifiers is to minimize the weighted cross entropy loss between the predicted and the true distributions:
\begin{equation}
    \mathcal{L} = -\sum_{i=1}^S (w^O y_i \log \hat{p_i} + w^R(1-y_i) \log (1-\hat{p_i})) + \lambda\left\|\Theta\right\|_2
\end{equation}
where $S$ denotes the number of training samples, $w^O$ is the weight of incorrectly predicting a rejected change, $w^R$ is the weight of incorrectly predicting an approved change. These two terms provide the opportunity to handle an imbalanced label distribution. $\lambda$ is the weight of the $L_2$ regularization term. $\Theta$ denotes all trainable parameters.

\section{Experimental design}
This section introduces the process of the experiment, including the repository selection and the data construction, baseline setting, evaluation metrics and experimental setting.

\begin{table}[t]
    \begin{center}
      \caption{Statistics of the AACR dataset.}
      \begin{tabular}{c|c|c|c}
        \hline
      
        \hline
        Repository & \#methods & \#rejected & reject rate \\
        \hline
        accumulo & 12,704 & 2,883 & 23\% \\
        ambari & 5,313 & 542 & 10\% \\
        cloudstack & 9,942 & 6,032 & 61\% \\
        commons-lang & 6,176 & 5634 & 91\% \\
        flink & 23,792 & 16,172 & 68\% \\
        incubator-point & 7,759 & 1,001 & 13\% \\
        kafka & 24,912 & 8,888 & 36\% \\
        lucene-solr & 6,785 & 2,886 & 43\% \\
        shardingsphere & 12,254 & 676 & 6\% \\
        \hline
        
        \hline
        \end{tabular}
    \end{center}
  \end{table}

\subsection{Dataset Construction}
We selected 9 projects from Github belonging to the Apache Foundation because the Apache Foundation is a widely used code review source. Six of them (\emph{commons-lang, flink, incubator-point, kafka, lucene-solr, shardingsphere}) were chosen because they have over 2000 stars. The remaining projects (\emph{accumulo, ambari, cloudstack}) were selected by \cite{shi2019automatic}. The language of all the projects is Java.

For data processing, we extracted all issues belonging to these projects from 2018 to 2020. Among these issues, many do not involve code submission, but only provide feedback, so we need to choose according to the issue type. After manually analyzing hundreds of issue types, we finally chose the types \emph{PullRequestEvent} and \emph{PullRequestReviewCommentEvent}. Because only these two types of issues have revised code, it can be judged whether the code has passed the review by inspecting whether the code has been added to the code base.

In many practical cases, we can easily extract the original code and revised code from the issue, but because these codes are usually contained in many files, it is difficult for us to use them directly as the input of the network. Therefore, we assume that all of the changes are independent and identically distributed, so there is no connection between these changes, and if a file contains many changed methods, we can split these methods independently as inputs. 

Further, if we add a new method or delete a whole method, half of the input data is empty. So we discard these data because they cannot be fed into the network. That is, we only consider the case where the code has been changed. In addition, considering that the submitted data may be too large, we subdivide the code submitted each time into the method level for subsequent experiments. 

After processing the data, each piece of data comprises three parts: the original code fragment, the revised code fragment, and the label. The original code fragment and the revised code fragment are both method-level Java programs. The label uses 0 and 1 to represent rejection and acceptance. The basic statistics of the AACR are summarized in Table 1.

In this paper, the rate of the rejection between 6\% and 91\% means that there is class imbalance during model training and it will lead to poor performance, so we set the 'class\_weight' parameter to 'balance' in our model.

\subsection{Comparison Models}
In this paper, we compared our proposed model (SimAST-GCN) with three other models in the ACR task. These models use different methods (including delimiter-based method, token-based method, tree-based method) to obtain the code features. The baseline models are as follows:

\begin{itemize}
\item \textbf{DACE} \cite{shi2019automatic} divides the code according to the delimiter and designs a pairwise recursive autoencoder to compare the code.
\item \textbf{TBRNN} serializes the AST into tokens and uses an RNN to capture the syntactic and semantic information.
\item \textbf{ASTNN} \cite{8812062} splits large ASTs into a sequence of small statement trees and calculates the representation distance to compare the code.
\end{itemize}

In addition, we considered variants of our proposed SimAST-GCN.

\begin{itemize}
    \item \textbf{ASTGCN} is our proposed model without the Simplified AST module.
    \item \textbf{SimAST} is our proposed model without the GCN module.
    \item \textbf{SAGCN-C} is our proposed model, but where it connect the representations to compare codes rather than calculating the distance.
    \end{itemize}

In order to ensure the fairness of the experiment and the stability of the results, we ran all the methods on the new AACR dataset, and each experiment was repeated 30 times.

\subsection{Evaluation}
\subsubsection{Metrics for ACR evaluation}
Since the automatic code review can be formulated as a binary classification problem (accept or reject) \cite{shi2019automatic}, we choose the commonly-used Accuracy, F1-measure (F1), Area under the receiver operating characteristic curve (AUC) as evaluation metrics. In addition, considering the unbalanced data distribution in the AACR dataset, we also added another evaluation metric Matthews correlation coefficient (MCC) to better evaluate the performance and efficiency of the model.

The value range of AUC is [0,1]. When the AUC value is 1, it means that the predicted value is consistent with the correct value. When the AUC value is 0.5, it is equivalent to random selection. When the AUC value is less than 0.5, it means that it is worse than random selection.

The detailed definitions of Accuracy is as follows:
\begin{equation}
	Accuracy=\frac{TP + TN}{TP + TN + FP + FN}
\end{equation}
where TP, FP, FN, and TN represent True Positives, False Positives, False Negatives, and True Negatives, respectively. The Accuracy is the ratio of the number of correctly predicted samples to the total number of predicted samples.

The calculation formula of F1 is as follows:
\begin{equation}
	Precision=\frac{TP}{TP + FP}
\end{equation}
\begin{equation}
	Recall=\frac{TP}{TP + FN}
\end{equation}
\begin{equation}
	F1=2*\frac{Precision*Recall}{Precision+Recall}
\end{equation}
The F1 score is the harmonic mean of the precision and recall. The value range of F1 is [0,1], which indicates the result is between the zero precision (or recall) value and the perfect recall and precision.

The calculation formula of MCC is:
\begin{small} 
\begin{equation}
	MCC = \frac{TP \times TN - TP \times FN}{\sqrt{(TP+FP)(TP+FN)(TN+FP)(TN+FN)}}
\end{equation}
\end{small}
The value range of MCC is [-1,1], indicating the prediction is totally wrong and the predicted result is consistent with the real situation, separately.

\subsubsection{Significance analysis (Win/Tie/Loss indicator)}
In this paper, we used the Win/Tie/Loss indicator to compare further the performance difference between SimAST-GCN and the baseline models, which is widely used in software fields \cite{8946058},\cite{8330212}. We repeated the experiment 30 times for all models. Then we applied two data analysis methods (Wilcoxon signed-rank test and Cliff's delta test) to analyze the performance of SimAST-GCN and other methods.

The Wilcoxon signed-rank test is commonly used for pairwise comparison. It is a non-parametric statistical hypothesis test used to determine whether the two populations of matched samples have the same distribution. Different from the Student's t-test, the Wilcoxon signed-rank test does not assume that the data are normally distributed. It is more statistically detectable for different datasets than the Student's t-test, and it is more likely to give statistically significant results.
The $p$ value is used to determine whether the difference between the two populations of a matched samples is significant ($p$ value $< 0.05$) or not.

The Cliff's delta test \cite{macbeth2011cliff} is a non-parametric effect size test, which is a supplementary analysis of the Wilcoxon signed-rank test in this paper. It measures the difference between two populations of comparison samples in the form of numerical value. Table 2 shows the mappings between Cliff's delta values ($|\delta|$) and their effective levels.

Specifically, for SimAST-GCN and a baseline model $M$, the Win/Tie/Loss indicator between them on ACR task is calculated as follows:

\begin{itemize}
    \item \textbf{Win}: The result of SimAST-GCN outperforms $M$, if the $p$ value less than 0.05, and the effective level of Cliff's delta is not $Negligible$.
    \item \textbf{Loss}: The result of $M$ outperforms SimAST-GCN, if the $p$ value less than 0.05, and the effective level of Cliff's delta is not $Negligible$.
    \item \textbf{Tie}: Others.
\end{itemize}

\subsection{Experimental Setting}
In our experiments, we used the javalang tools\footnote{https://github.com/c2nes/javalang} to obtain ASTs for Java code, and we used Skip-gram algorithm implemented by gensim library \cite{rehurek_lrec} to train the embeddings of nodes. The embedding size was set to 300. The hidden size of Bi-GRU was 300. The number of GCN layers was 3, which is the optimal depth in pilot studies. The coefficients $w^O$ and $w^R$ were related to the dataset, and the coefficient $\lambda$ of $L_2$ regularization item was set to $10^{-5}$. Adam was utilized as the optimizer with a learning rate of $10^{-3}$ to train the model, and the mini-batch was 128. We random initialized all the $\mathbf{W}$ and $\mathbf{b}$ with a uniform distribution.

All the experiments were conducted on a server with 24 cores of 3.8GHz CPU and a NVIDIA GeForce RTX 3090 GPU.

\begin{table}[htbp]
	\footnotesize
	\caption{Mappings between the Cliff's delta values($|\delta|$) and their effective levels}
	\label{tab_delta}
	\tabcolsep 30pt
	\centering
	\begin{tabular}{ ll}
		\toprule
		Cliff's delta                  & Effective levels \\\midrule
		$|\delta| <$ 0.147             & Negligible       \\
		0.147 $\leq |\delta| <$ 0.33   & Small            \\
		0.33 $\leq |\delta| <$ 0.474   & Medium           \\
		0.474 $\leq |\delta|$         & Large             \\
		\bottomrule
	\end{tabular}
\end{table}

\begin{table*}[h!]
\begin{center}
\caption{Accuracy values of SimAST-GCN and other three baseline methods.}
\resizebox{\textwidth}{!}{
\begin{tabular}{c | c c c c | c c c}
\hline
 
\hline
\multirow{2}{*}{Repository} & \multicolumn{4}{c}{Accuracy} & \multicolumn{3}{c}{p($\delta$)}\\
\cline{2-8}
& DACE & TBRNN & ASTNN & SimAST-GCN & DACE vs. SimAST-GCN & TBRNN vs. SimAST-GCN & ASTNN vs. SimAST-GCN\\
\hline
accumulo & 86.108 & 88.868 & 90.887 & \textbf{94.597} & <0.05(+Large) & <0.05(+Large) & <0.05(+Large)\\
ambari & 75.563 & 70.944 & 82.015 & \textbf{85.969} & <0.05(+Large) & <0.05(+Large) & <0.05(+Medium)\\
cloudstack & 74.181 & 75.957 & 72.311 & \textbf{76.664} & <0.05(+Large) & <0.05(+Large) & <0.05(+Large)\\
commons-lang & 96.525 & 97.758 & 93.381 & \textbf{98.306} & <0.05(+Large) & <0.05(+Large) & <0.05(+Large)\\
flink & 67.133 & \textbf{70.969} & 69.438 & 70.125 & <0.05(+Large) & <0.05(-Large) & <0.05(+Medium)\\
incubator-pinot & 75.168 & 78.605 & 61.605 & \textbf{78.888} & <0.05(+Small) & 0.6(-Small) & <0.05(+Large)\\
kafka & 57.71 & 58.885 & \textbf{63.74} & 62.669 & <0.05(+Large) & <0.05(+Large) & <0.05(-Medium)\\
lucene-solr & 67.882 & 72.95 & 71.45 & \textbf{74.0} & <0.05(+Large) & <0.05(+Large) & <0.05(+Large)\\
shardingsphere & 86.994 & 88.315 & \textbf{89.975} & 89.695 & <0.05(+Large) & <0.05(+Medium) & <0.05(-Large)\\
\hline
Average \& & \multirow{2}{*}{76.363} & \multirow{2}{*}{78.139} & \multirow{2}{*}{77.2} &\multirow{2}{*}{\textbf{81.213}} &\multirow{2}{*}{9/0/0} &\multirow{2}{*}{7/1/1} &\multirow{2}{*}{7/0/2} \\
Win/Tie/Loss & & & & & & &\\
 
\hline

\hline
\end{tabular}
}
\end{center}
\end{table*}
\begin{table*}[h!]
\begin{center}
\caption{F1 values of SimAST-GCN and other three baseline methods.}
\resizebox{\textwidth}{!}{
\begin{tabular}{c | c c c c | c c c}
\hline
 
\hline
\multirow{2}{*}{Repository} & \multicolumn{4}{c}{F1} & \multicolumn{3}{c}{p($\delta$)}\\
\cline{2-8}
& DACE & TBRNN & ASTNN & SimAST-GCN & DACE vs. SimAST-GCN & TBRNN vs. SimAST-GCN & ASTNN vs. SimAST-GCN\\
\hline
accumulo & 0.91 & 0.926 & 0.942 & \textbf{0.965} & <0.05(+Large) & <0.05(+Large) & <0.05(+Large)\\
ambari & 0.855 & 0.82 & 0.896 & \textbf{0.921} & <0.05(+Large) & <0.05(+Large) & <0.05(+Medium)\\
cloudstack & 0.683 & 0.727 & 0.726 & \textbf{0.729} & <0.05(+Large) & <0.05(+Large) & <0.05(+Large)\\
commons-lang & 0.794 & 0.873 & 0.679 & \textbf{0.9} & <0.05(+Large) & <0.05(+Large) & <0.05(+Large)\\
flink & 0.554 & 0.588 & 0.559 & \textbf{0.598} & <0.05(+Large) & <0.05(+Large) & <0.05(+Large)\\
incubator-pinot & 0.85 & 0.873 & 0.732 & \textbf{0.874} & <0.05(+Small) & 0.6(-Small) & <0.05(+Large)\\
kafka & 0.623 & 0.625 & 0.698 & \textbf{0.699} & <0.05(+Large) & <0.05(+Large) & 0.417(-Negligible)\\
lucene-solr & 0.731 & 0.749 & 0.752 & \textbf{0.771} & <0.05(+Large) & <0.05(+Large) & <0.05(+Large)\\
shardingsphere & 0.929 & 0.936 & \textbf{0.945} & 0.944 & <0.05(+Large) & <0.05(+Medium) & <0.05(-Large)\\
\hline
Average \& & \multirow{2}{*}{0.77} & \multirow{2}{*}{0.791} & \multirow{2}{*}{0.77} &\multirow{2}{*}{\textbf{0.822}} &\multirow{2}{*}{9/0/0} &\multirow{2}{*}{8/1/0} &\multirow{2}{*}{7/1/1} \\
Win/Tie/Loss & & & & & & &\\
 
\hline

\hline
\end{tabular}
}
\end{center}
\end{table*}
\begin{table*}[h!]
\begin{center}
\caption{AUC values of SimAST-GCN and other three baseline methods.}
\resizebox{\textwidth}{!}{
\begin{tabular}{c | c c c c | c c c}
\hline
 
\hline
\multirow{2}{*}{Repository} & \multicolumn{4}{c}{AUC} & \multicolumn{3}{c}{p($\delta$)}\\
\cline{2-8}
& DACE & TBRNN & ASTNN & SimAST-GCN & DACE vs. SimAST-GCN & TBRNN vs. SimAST-GCN & ASTNN vs. SimAST-GCN\\
\hline
accumulo & 0.805 & 0.877 & 0.869 & \textbf{0.925} & <0.05(+Large) & <0.05(+Large) & <0.05(+Large)\\
ambari & 0.653 & 0.605 & 0.651 & \textbf{0.668} & <0.05(+Medium) & <0.05(+Large) & <0.05(+Large)\\
cloudstack & 0.734 & 0.768 & 0.762 & \textbf{0.768} & <0.05(+Large) & 0.136(+Negligible) & <0.05(+Large)\\
commons-lang & 0.9 & 0.962 & 0.956 & \textbf{0.971} & <0.05(+Large) & <0.05(+Large) & <0.05(+Large)\\
flink & 0.671 & 0.695 & 0.678 & \textbf{0.711} & <0.05(+Large) & <0.05(+Large) & <0.05(+Large)\\
incubator-pinot & 0.622 & 0.623 & 0.652 & \textbf{0.656} & <0.05(+Large) & <0.05(+Large) & <0.05(+Small)\\
kafka & 0.602 & 0.612 & 0.638 & \textbf{0.644} & <0.05(+Large) & <0.05(+Large) & <0.05(+Large)\\
lucene-solr & 0.689 & 0.737 & 0.709 & \textbf{0.745} & <0.05(+Large) & <0.05(+Large) & <0.05(+Large)\\
shardingsphere & 0.684 & 0.745 & 0.783 & \textbf{0.785} & <0.05(+Large) & <0.05(+Large) & 0.491(+Negligible)\\
\hline
Average \& & \multirow{2}{*}{0.707} & \multirow{2}{*}{0.736} & \multirow{2}{*}{0.744} &\multirow{2}{*}{\textbf{0.764}} &\multirow{2}{*}{9/0/0} &\multirow{2}{*}{8/1/0} &\multirow{2}{*}{8/1/0} \\
Win/Tie/Loss & & & & & & &\\
 
\hline

\hline
\end{tabular}
}
\end{center}
\end{table*}
\begin{table*}[h!]
\begin{center}
\caption{MCC values of SimAST-GCN and other three baseline methods.}
\resizebox{\textwidth}{!}{
\begin{tabular}{c | c c c c | c c c}
\hline
 
\hline
\multirow{2}{*}{Repository} & \multicolumn{4}{c}{MCC} & \multicolumn{3}{c}{p($\delta$)}\\
\cline{2-8}
& DACE & TBRNN & ASTNN & SimAST-GCN & DACE vs. SimAST-GCN & TBRNN vs. SimAST-GCN & ASTNN vs. SimAST-GCN\\
\hline
accumulo & 0.61 & 0.709 & 0.735 & \textbf{0.847} & <0.05(+Large) & <0.05(+Large) & <0.05(+Large)\\
ambari & 0.213 & 0.141 & 0.239 & \textbf{0.246} & <0.05(+Large) & <0.05(+Large) & <0.05(+Small)\\
cloudstack & 0.469 & 0.527 & 0.521 & \textbf{0.533} & <0.05(+Large) & <0.05(+Large) & <0.05(+Large)\\
commons-lang & 0.775 & 0.863 & 0.687 & \textbf{0.892} & <0.05(+Large) & <0.05(+Large) & <0.05(+Large)\\
flink & 0.32 & 0.373 & 0.337 & \textbf{0.391} & <0.05(+Large) & <0.05(+Large) & <0.05(+Large)\\
incubator-pinot & 0.182 & 0.204 & 0.207 & \textbf{0.222} & <0.05(+Large) & <0.05(+Large) & <0.05(+Large)\\
kafka & 0.202 & 0.218 & 0.265 & \textbf{0.276} & <0.05(+Large) & <0.05(+Large) & <0.05(+Large)\\
lucene-solr & 0.376 & 0.469 & 0.423 & \textbf{0.485} & <0.05(+Large) & <0.05(+Large) & <0.05(+Large)\\
shardingsphere & 0.203 & 0.36 & \textbf{0.393} & 0.369 & <0.05(+Large) & <0.05(+Medium) & <0.05(-Large)\\
\hline
Average \& & \multirow{2}{*}{0.372} & \multirow{2}{*}{0.429} & \multirow{2}{*}{0.423} &\multirow{2}{*}{\textbf{0.474}} &\multirow{2}{*}{9/0/0} &\multirow{2}{*}{9/0/0} &\multirow{2}{*}{8/0/1} \\
Win/Tie/Loss & & & & & & &\\
 
\hline

\hline
\end{tabular}
}
\end{center}
\end{table*}

\begin{figure*}[ht]
    \begin{center}
        \includegraphics[width=0.49\linewidth]{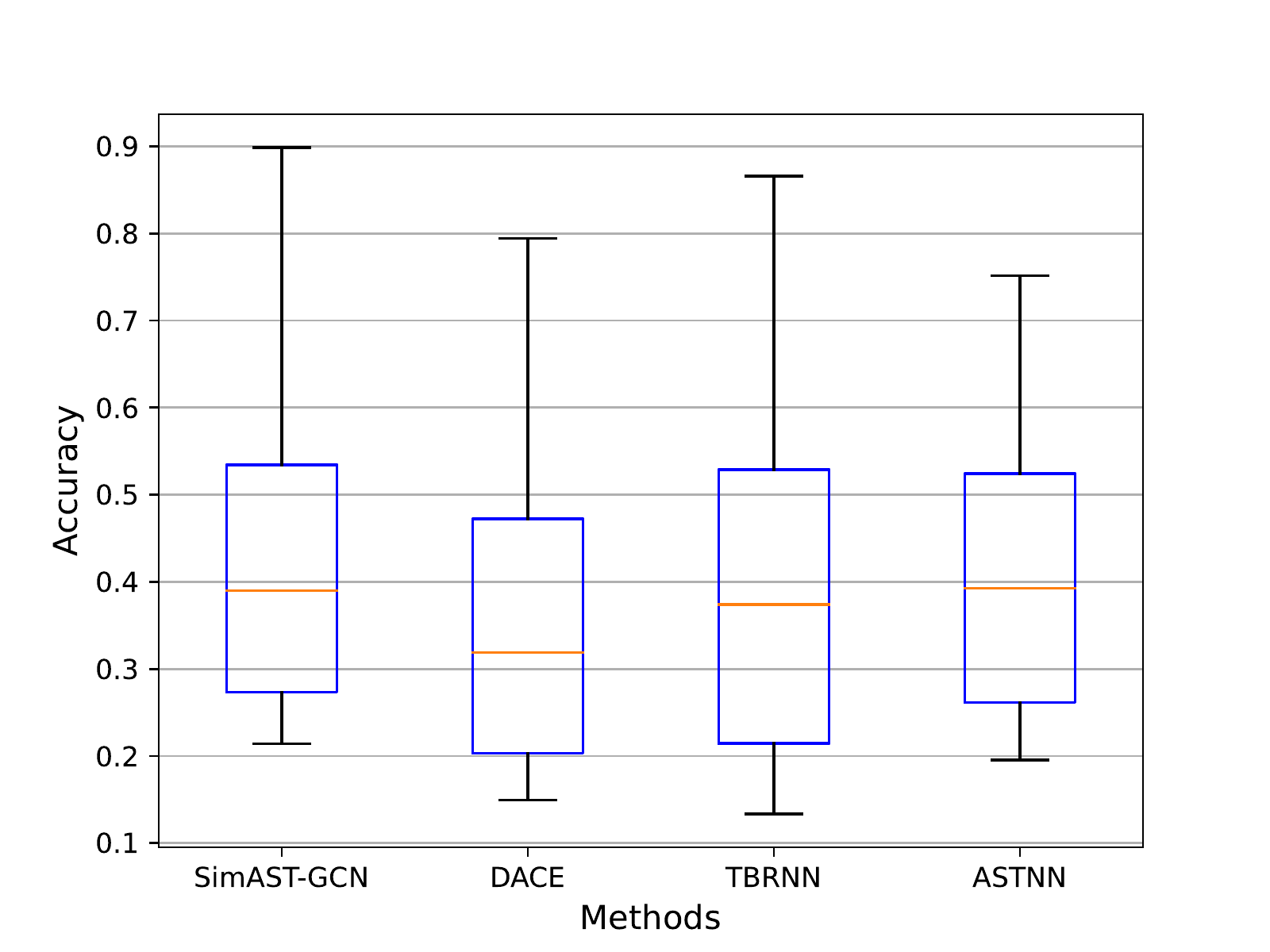}  
        \includegraphics[width=0.49\linewidth]{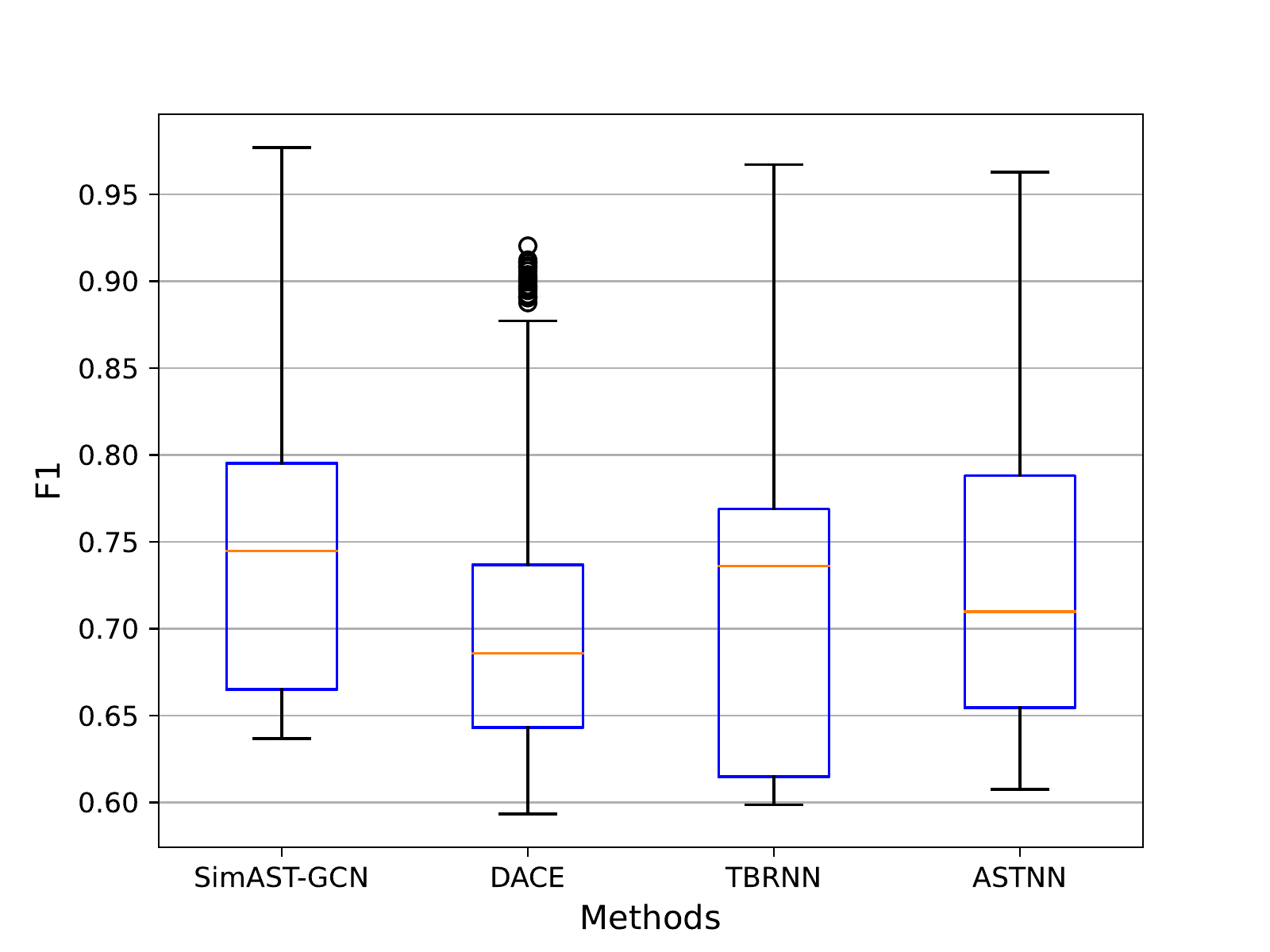}  

        \includegraphics[width=0.49\linewidth]{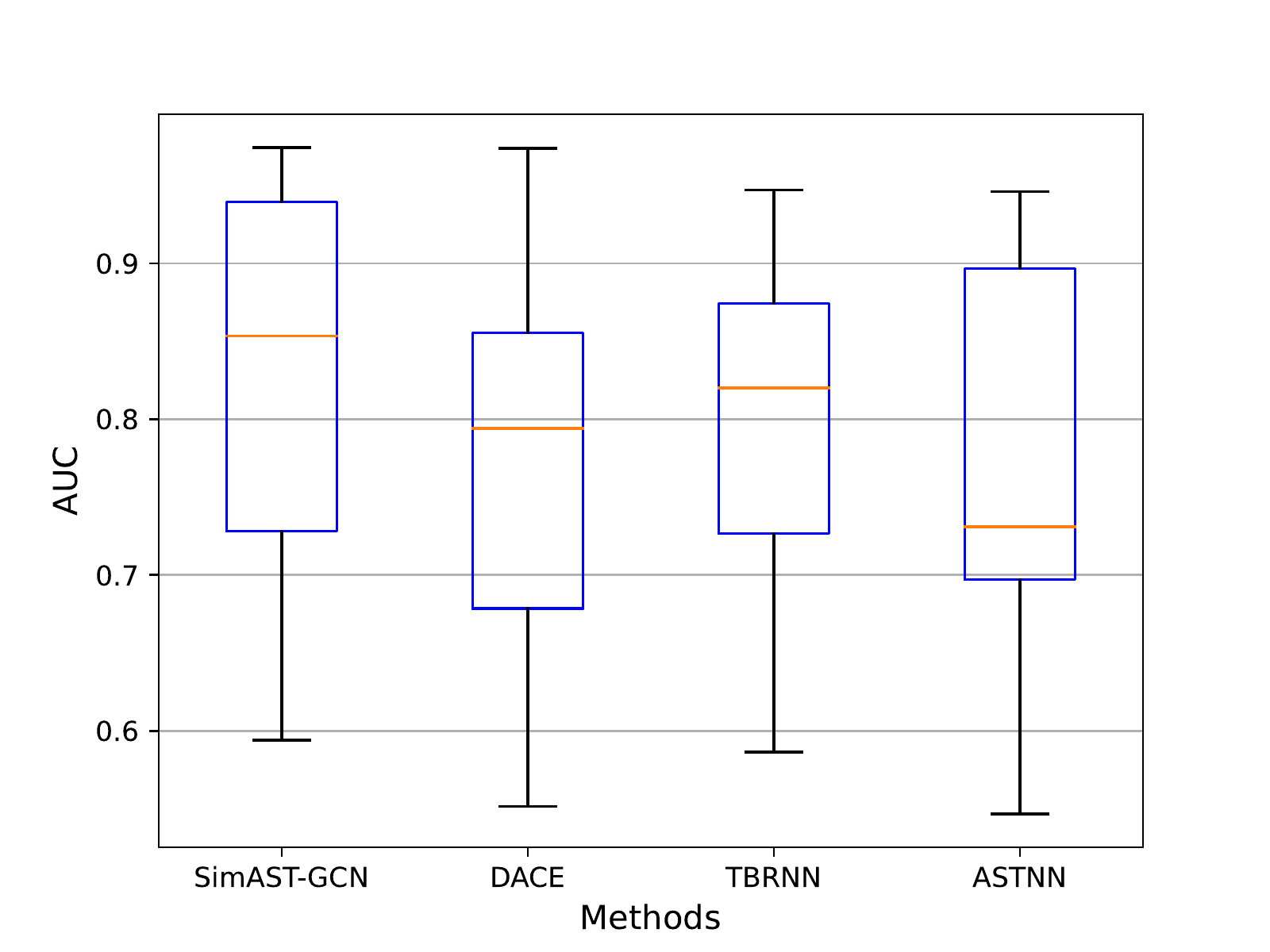} 
        \includegraphics[width=0.49\linewidth]{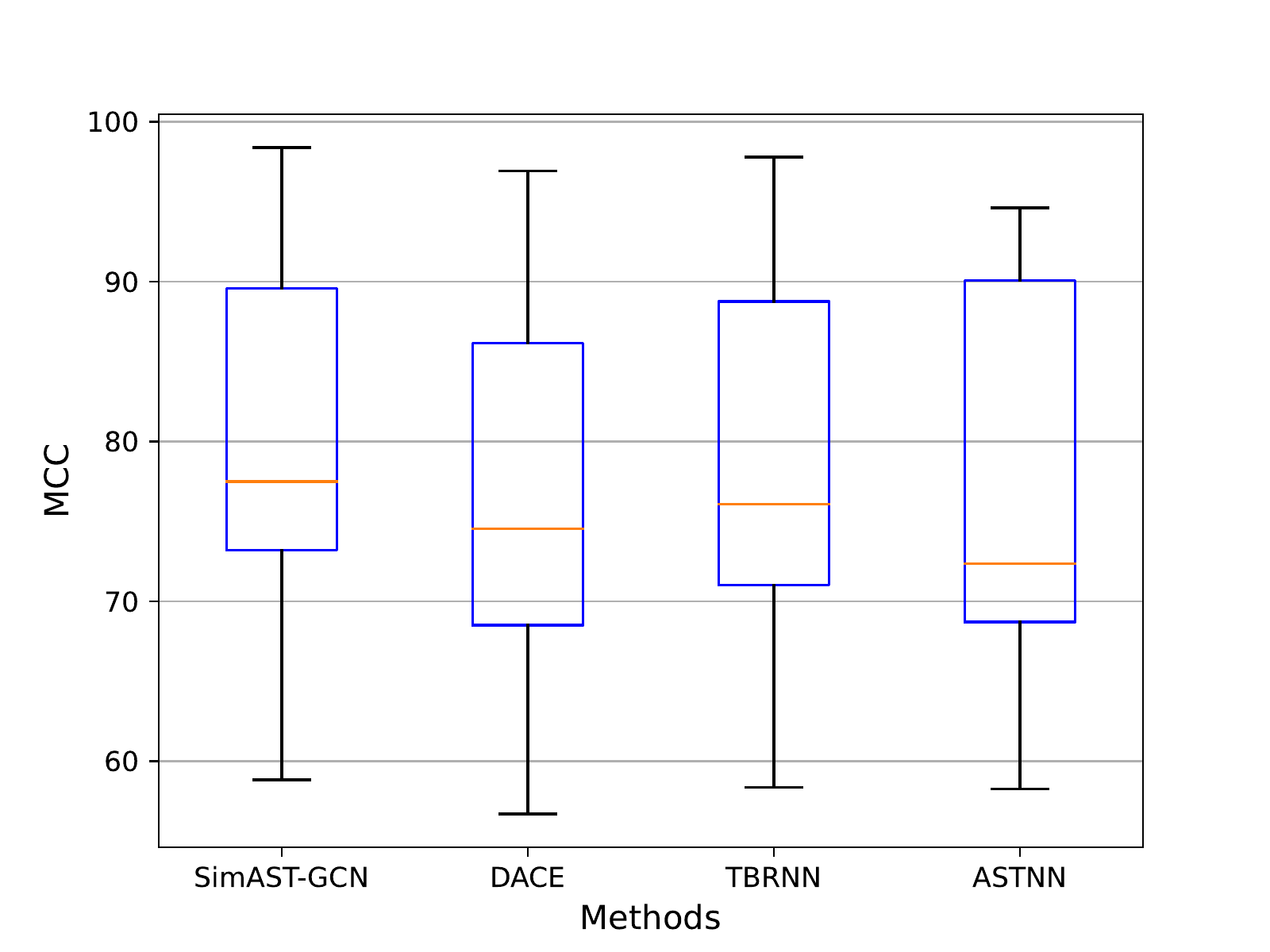} 
        \caption{Boxplot of four metrics of SimAST-GCN and the other three baseline methods.}
    \end{center}  
\end{figure*}

\section{Experimental Results}
This section shows the performance of our proposed method SimAST-GCN with other baseline methods. Therefore, we put forward the following research questions: 
\begin{itemize}
	\item RQ1: Does our proposed SimAST-GCN model outperform other models for automatic code review?
	\item RQ2: How different parameter settings and module influence the performance of our method?
	\item RQ3: Does our proposed SimAST-GCN model outperform other models in terms of time efficiency?
\end{itemize}


\subsection{Does our proposed SimAST-GCN model outperform other models for automatic code review?}

Tables 3--6 show the comparison results of four metrics on the AACR dataset. For each table, it is divided into three columns. The first column is the repository name. The second column is the corresponding metric column, which shows the metric values of our proposed SimAST-GCN method and the other three baseline methods. For each repository, the result of the best method is presented in hold. The $p$($\delta$) column shows the $p$ values of Wilcoxon signed-rank test and Cliff's delta values between Sim-AST and the other three baseline methods. For the $p$ value of Wilcoxon signed-rank test, we displayed the original value in the table if the value is not less than 0.05. Otherwise, we will display '$<0.05$' in the table. For Cliff's delta value, we displayed the effective level (shown in Table 2) in the table. In order to distinguish the positive and negative Cliff's delta value, we used '+' and '-' before the effective level to represent the property. The row 'Average \& Win/Tie/Loss' shows the average value of the corresponding metric and the Win/Tie/Loss indicator.

The results in all the metrics show that the proposed SimAST-GCN consistently outperforms all comparison models. This verifies the effectiveness of our proposed method at ACR. 

Compared with the delimiter-based model (DACE), we find that SimAST-GCN achieves the best performance in all the terms of the F1, AUC, and MCC. This is because, compared to DACE, SimAST-GCN is not the delimiter-based method. We adopt an AST as the abstract code representation, which demonstrates the effectiveness of AST at code representation.

Compared with the AST-based models (TBRNN and ASTNN), SimAST-GCN also achieves the best performance according to all metrics. Although both TBRNN and ASTNN use an abstract syntax tree for source code processing, in the code review task, we find that the two methods do not perform well in comparison to SimAST-GCN. On the one hand, we simplify the AST and enhance the connection properties between nodes. On the other hand, we use a more advanced graph neural network to model the simplified AST to capture better syntactic and semantic information. 

Moreover, we utilize a retrieval-based attention mechanism to capture significant sentiment features from the context representations for the source code. This mechanism dramatically improves the performance of the model, allowing the model to focus on the parts of the code that may have changed, which is widely used in natural language processing.

In summary, this is the first study to simplify the abstract syntax tree and deploy a graph structure to leverage Simplified AST for the code review task. Our experiments show the efficiency of exploiting the Simplified AST and adopting the graph neural network.

\subsection{How different parameter settings and module influence the performance of our method?}

\begin{figure*}[ht]
    \begin{center}
        \includegraphics[width=0.49\linewidth]{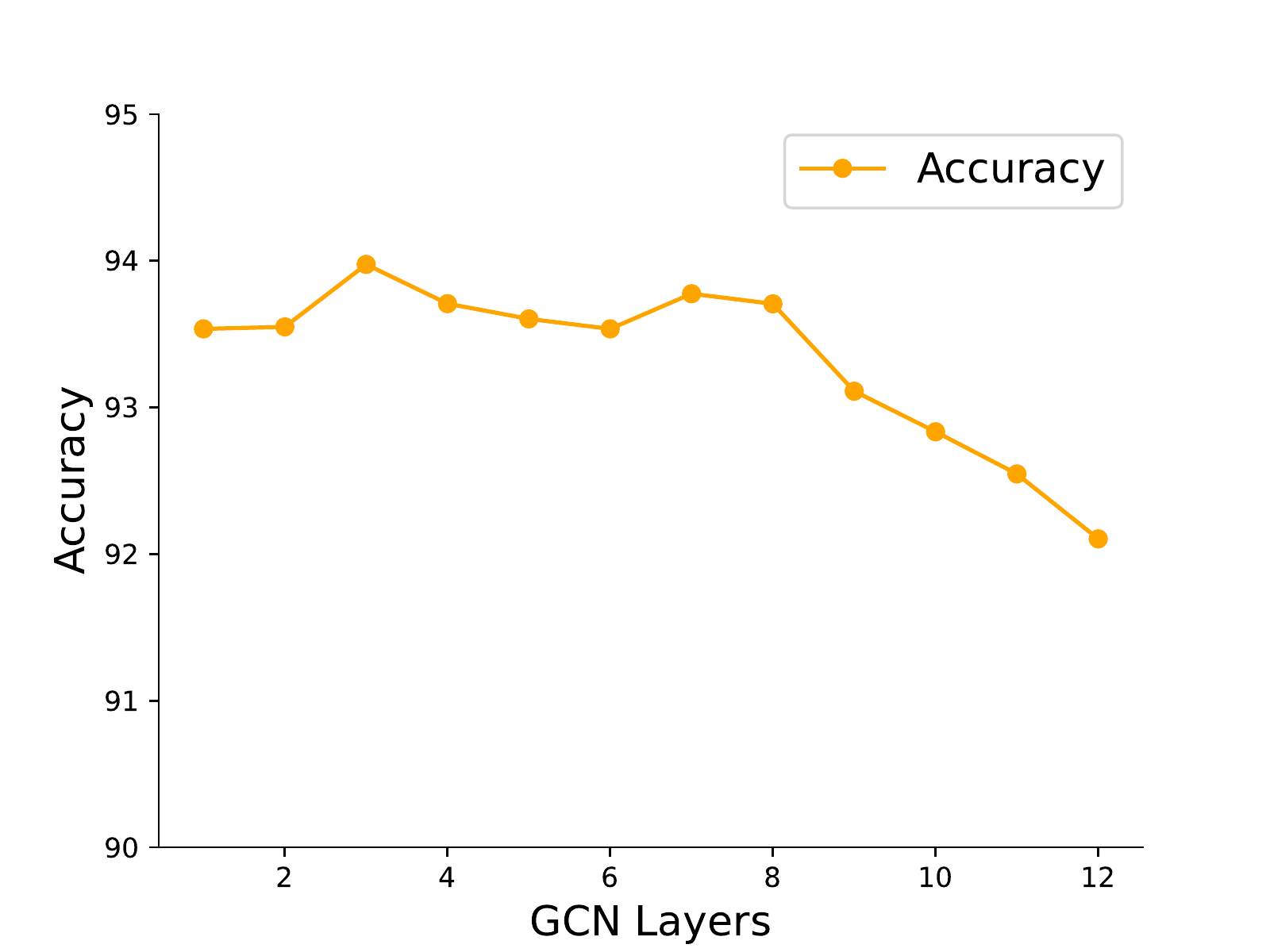}  
        \includegraphics[width=0.49\linewidth]{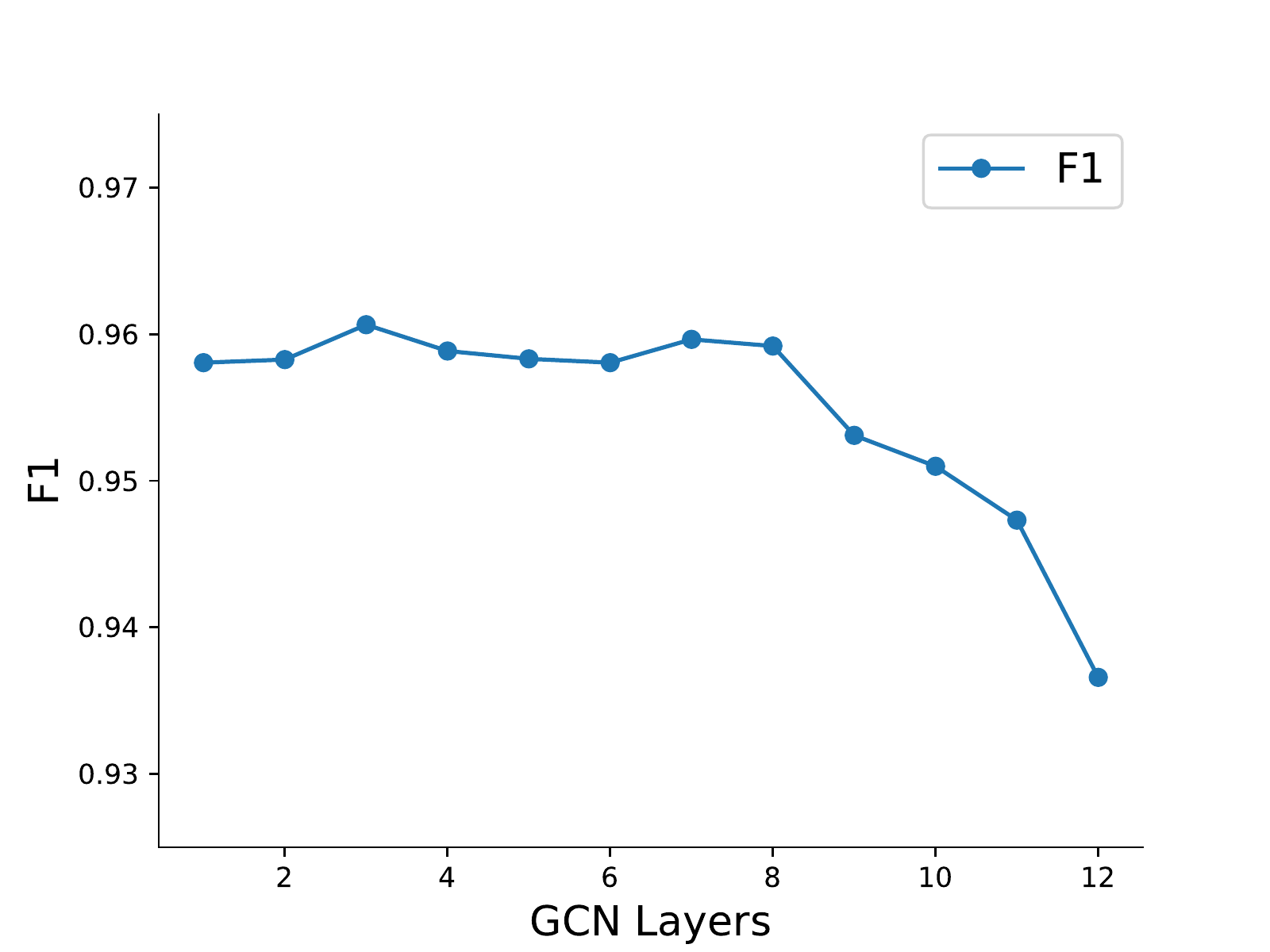}  

        \includegraphics[width=0.49\linewidth]{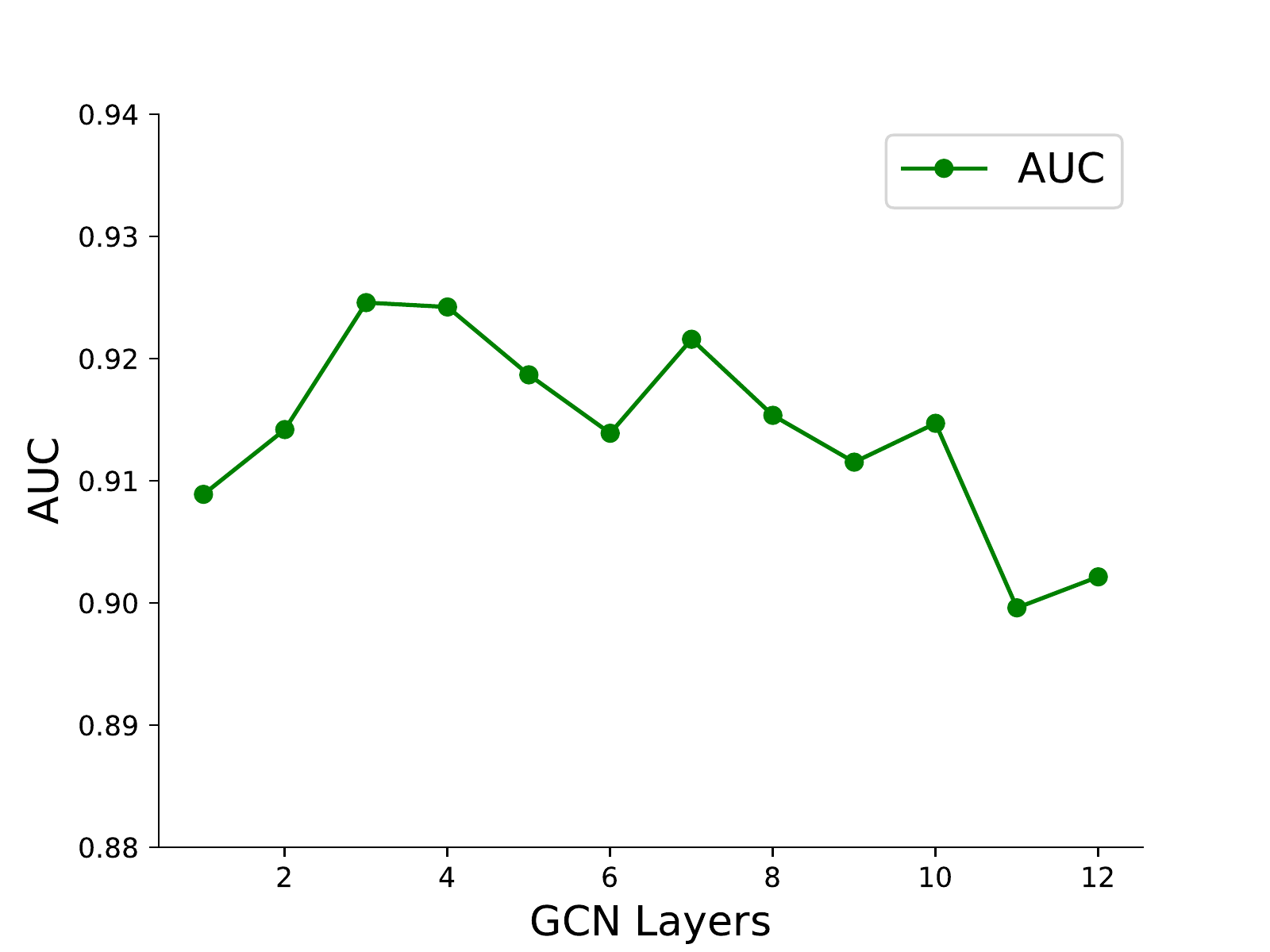}  
        \includegraphics[width=0.49\linewidth]{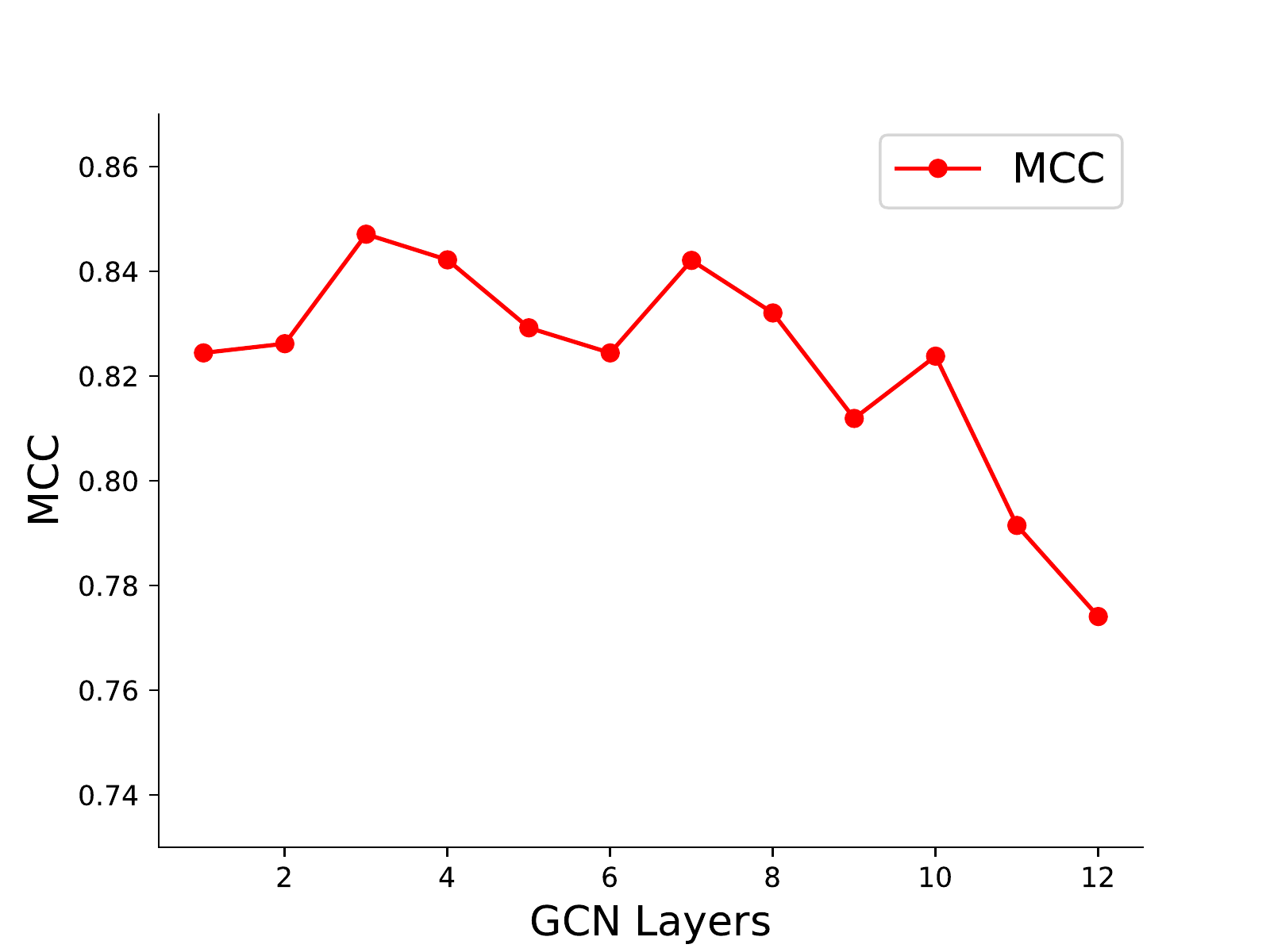} 
        \caption{Impact of the number of GCN layers. Four metrics based on different numbers of GCN layers are reported.}
    \end{center}  
\end{figure*}

\subsubsection{Parameter settings}
As a key component of our model, we investigated the impact of the GCN layer number on the performance of our proposed method SimAST-GCN. We varied the number of layers from 1 to 12, and we report the results of four metrics in Figure 6. Overall, the 3-layer GCN achieves the best performance on the \emph{accumulo} dataset. Hence, we finally set the number of GCN layers to 3 in our experiments. 

Comparatively, the layers of GCN less than 3 in SimAST-GCN perform unsatisfactorily, which indicates that fewer GCN layers in SimAST-GCN are insufficient to derive the precise syntactical dependencies of the source code.

In addition, the performance of SimAST-GCN decreases as the number of GCN layers increases, and tends to decrease when the depth of the model is greater than 7. This means that simply increasing the depth of the GCN will reduce the learning ability of the model because the model parameters increase sharply.


\begin{table*}[h!]
\begin{center}
\caption{Accuracy values of SimAST-GCN and other three variant methods.}
\resizebox{\textwidth}{!}{
\begin{tabular}{c | c c c c | c c c}
\hline
 
\hline
\multirow{2}{*}{Repository} & \multicolumn{4}{c}{Accuracy} & \multicolumn{3}{c}{p($\delta$)}\\
\cline{2-8}
& ASTGCN & SimAST & SAGCN-C & SimAST-GCN & ASTGCN vs. SimAST-GCN & SimAST vs. SimAST-GCN & SAGCN-C vs. SimAST-GCN\\
\hline
accumulo & 84.53 & 93.567 & 92.476 & \textbf{94.597} & <0.05(+Large) & <0.05(+Large) & <0.05(+Large)\\
ambari & 80.96 & 78.813 & 75.102 & \textbf{85.969} & <0.05(+Large) & <0.05(+Large) & <0.05(+Large)\\
cloudstack & 71.54 & 76.507 & 71.725 & \textbf{76.664} & <0.05(+Large) & 0.719(+Negligible) & <0.05(+Large)\\
commons-lang & 97.588 & 98.213 & 96.861 & \textbf{98.306} & <0.05(+Large) & 0.146(-Medium) & <0.05(+Large)\\
flink & 66.634 & 69.552 & 67.493 & \textbf{70.125} & <0.05(+Large) & <0.05(+Large) & <0.05(+Large)\\
incubator-pinot & 70.004 & 73.961 & 74.319 & \textbf{78.888} & <0.05(+Large) & <0.05(+Medium) & <0.05(+Medium)\\
kafka & \textbf{64.115} & 61.845 & 59.946 & 62.669 & <0.05(-Large) & <0.05(+Medium) & <0.05(+Large)\\
lucene-solr & 58.08 & 72.953 & 73.171 & \textbf{74.0} & <0.05(+Large) & <0.05(+Large) & <0.05(+Large)\\
shardingsphere & 89.181 & 88.849 & 87.315 & \textbf{89.695} & <0.05(+Small) & <0.05(+Small) & <0.05(+Large)\\
\hline
Average \& & \multirow{2}{*}{75.848} & \multirow{2}{*}{79.362} & \multirow{2}{*}{77.601} &\multirow{2}{*}{\textbf{81.213}} &\multirow{2}{*}{8/0/1} &\multirow{2}{*}{7/2/0} &\multirow{2}{*}{9/0/0} \\
Win/Tie/Loss & & & & & & &\\
 
\hline

\hline
\end{tabular}
}
\end{center}
\end{table*}
\begin{table*}[h!]
\begin{center}
\caption{F1 values of SimAST-GCN and other three variant methods.}
\resizebox{\textwidth}{!}{
\begin{tabular}{c | c c c c | c c c}
\hline
 
\hline
\multirow{2}{*}{Repository} & \multicolumn{4}{c}{F1} & \multicolumn{3}{c}{p($\delta$)}\\
\cline{2-8}
& ASTGCN & SimAST & SAGCN-C & SimAST-GCN & ASTGCN vs. SimAST-GCN & SimAST vs. SimAST-GCN & SAGCN-C vs. SimAST-GCN\\
\hline
accumulo & 0.897 & 0.959 & 0.951 & \textbf{0.965} & <0.05(+Large) & <0.05(+Large) & <0.05(+Large)\\
ambari & 0.889 & 0.876 & 0.85 & \textbf{0.921} & <0.05(+Large) & <0.05(+Large) & <0.05(+Large)\\
cloudstack & 0.72 & 0.717 & 0.688 & \textbf{0.729} & <0.05(+Large) & <0.05(+Large) & <0.05(+Large)\\
commons-lang & 0.859 & \textbf{0.902} & 0.815 & 0.9 & <0.05(+Large) & 0.06(-Large) & <0.05(+Large)\\
flink & 0.572 & 0.58 & 0.565 & \textbf{0.598} & <0.05(+Large) & <0.05(+Large) & <0.05(+Large)\\
incubator-pinot & 0.808 & 0.837 & 0.843 & \textbf{0.874} & <0.05(+Large) & <0.05(+Medium) & <0.05(+Medium)\\
kafka & \textbf{0.714} & 0.678 & 0.656 & 0.699 & <0.05(-Medium) & <0.05(+Large) & <0.05(+Large)\\
lucene-solr & 0.733 & 0.758 & 0.765 & \textbf{0.771} & <0.05(+Large) & <0.05(+Large) & <0.05(+Large)\\
shardingsphere & 0.941 & 0.939 & 0.931 & \textbf{0.944} & <0.05(+Small) & 0.094(+Negligible) & <0.05(+Large)\\
\hline
Average \& & \multirow{2}{*}{0.792} & \multirow{2}{*}{0.805} & \multirow{2}{*}{0.785} &\multirow{2}{*}{\textbf{0.822}} &\multirow{2}{*}{8/0/1} &\multirow{2}{*}{7/2/0} &\multirow{2}{*}{9/0/0} \\
Win/Tie/Loss & & & & & & &\\
 
\hline

\hline
\end{tabular}
}
\end{center}
\end{table*}
\begin{table*}[h!]
\begin{center}
\caption{AUC values of SimAST-GCN and other three variant methods.}
\resizebox{\textwidth}{!}{
\begin{tabular}{c | c c c c | c c c}
\hline
 
\hline
\multirow{2}{*}{Repository} & \multicolumn{4}{c}{AUC} & \multicolumn{3}{c}{p($\delta$)}\\
\cline{2-8}
& ASTGCN & SimAST & SAGCN-C & SimAST-GCN & ASTGCN vs. SimAST-GCN & SimAST vs. SimAST-GCN & SAGCN-C vs. SimAST-GCN\\
\hline
accumulo & 0.846 & 0.91 & 0.893 & \textbf{0.925} & <0.05(+Large) & <0.05(+Large) & <0.05(+Large)\\
ambari & \textbf{0.691} & 0.62 & 0.657 & 0.668 & <0.05(-Large) & <0.05(+Large) & <0.05(+Large)\\
cloudstack & 0.742 & 0.764 & 0.726 & \textbf{0.768} & <0.05(+Large) & <0.05(+Large) & <0.05(+Large)\\
commons-lang & 0.945 & \textbf{0.975} & 0.925 & 0.971 & <0.05(+Large) & 0.09(-Small) & <0.05(+Large)\\
flink & 0.689 & 0.686 & 0.682 & \textbf{0.711} & <0.05(+Large) & <0.05(+Large) & <0.05(+Large)\\
incubator-pinot & 0.662 & 0.651 & \textbf{0.668} & 0.656 & <0.05(-Medium) & <0.05(+Medium) & <0.05(-Large)\\
kafka & \textbf{0.65} & 0.616 & 0.619 & 0.644 & <0.05(-Large) & <0.05(+Large) & <0.05(+Large)\\
lucene-solr & 0.506 & 0.732 & 0.737 & \textbf{0.745} & <0.05(+Large) & <0.05(+Large) & <0.05(+Large)\\
shardingsphere & 0.747 & 0.686 & 0.703 & \textbf{0.785} & <0.05(+Large) & <0.05(+Large) & <0.05(+Large)\\
\hline
Average \& & \multirow{2}{*}{0.72} & \multirow{2}{*}{0.738} & \multirow{2}{*}{0.734} &\multirow{2}{*}{\textbf{0.764}} &\multirow{2}{*}{6/0/3} &\multirow{2}{*}{8/1/0} &\multirow{2}{*}{8/0/1} \\
Win/Tie/Loss & & & & & & &\\
 
\hline

\hline
\end{tabular}
}
\end{center}
\end{table*}
\begin{table*}[h!]
\begin{center}
\caption{MCC values of SimAST-GCN and other three variant methods.}
\resizebox{\textwidth}{!}{
\begin{tabular}{c | c c c c | c c c}
\hline
 
\hline
\multirow{2}{*}{Repository} & \multicolumn{4}{c}{MCC} & \multicolumn{3}{c}{p($\delta$)}\\
\cline{2-8}
& ASTGCN & SimAST & SAGCN-C & SimAST-GCN & ASTGCN vs. SimAST-GCN & SimAST vs. SimAST-GCN & SAGCN-C vs. SimAST-GCN\\
\hline
accumulo & 0.627 & 0.814 & 0.786 & \textbf{0.847} & <0.05(+Large) & <0.05(+Large) & <0.05(+Large)\\
ambari & \textbf{0.275} & 0.185 & 0.211 & 0.246 & <0.05(-Large) & <0.05(+Large) & <0.05(+Large)\\
cloudstack & 0.488 & 0.521 & 0.447 & \textbf{0.533} & <0.05(+Large) & <0.05(+Large) & <0.05(+Large)\\
commons-lang & 0.847 & \textbf{0.896} & 0.801 & 0.892 & <0.05(+Large) & 0.057(-Large) & <0.05(+Large)\\
flink & 0.349 & 0.352 & 0.337 & \textbf{0.391} & <0.05(+Large) & <0.05(+Large) & <0.05(+Large)\\
incubator-pinot & 0.221 & 0.229 & \textbf{0.24} & 0.222 & 0.813(+Negligible) & <0.05(-Large) & <0.05(-Large)\\
kafka & \textbf{0.288} & 0.222 & 0.228 & 0.276 & <0.05(-Large) & <0.05(+Large) & <0.05(+Large)\\
lucene-solr & 0.046 & 0.457 & 0.469 & \textbf{0.485} & <0.05(+Large) & <0.05(+Large) & <0.05(+Large)\\
shardingsphere & 0.328 & 0.263 & 0.231 & \textbf{0.369} & <0.05(+Large) & <0.05(+Large) & <0.05(+Large)\\
\hline
Average \& & \multirow{2}{*}{0.385} & \multirow{2}{*}{0.438} & \multirow{2}{*}{0.417} &\multirow{2}{*}{\textbf{0.474}} &\multirow{2}{*}{6/1/2} &\multirow{2}{*}{7/1/1} &\multirow{2}{*}{8/0/1} \\
Win/Tie/Loss & & & & & & &\\
 
\hline

\hline
\end{tabular}
}
\end{center}
\end{table*}

\subsubsection{Module influence}
We conducted an ablation study to analyze further the impact of different components of the proposed SimAST-GCN. The results of the four metrics are shown in Table 7--10. 

We can observe that the model without the Simplified AST (ASTGCN) performs most unsatisfactorily on the AACR dataset. This confirms that simplifying the AST is the most significant improvement for ACR. 

In addition, the removal of retrieval-based attention and GCN (SimAST) leads to a considerable performance drop. This indicates that the attention mechanism and node relations vastly improve the performance of ACR. 

We further observe that the model with the connected information (SAGCN-C) declines sharply, which indicates that it is better to calculate the distance between the representations rather than connecting the representation.




\begin{figure}[ht]
    \begin{center}
        
        \includegraphics[width=0.98\linewidth]{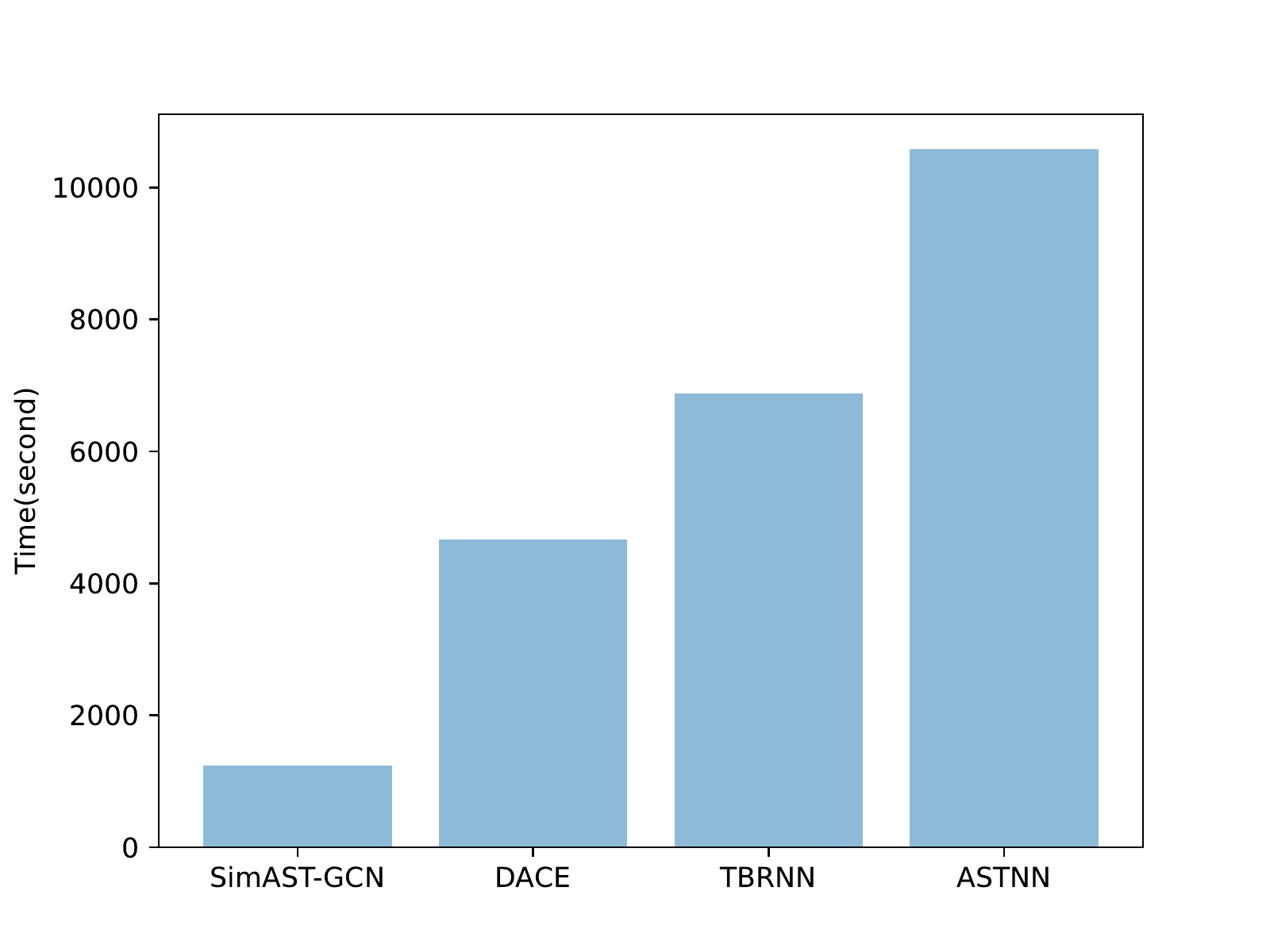} 
        \caption{Time consumption for SimAST-GCN and three baseline methods during training phrase.}
    \end{center}  
\end{figure}

\subsection{Does our proposed SimAST-GCN model outperform other models in terms of time efficiency?}
In practical applications, the training time of the model has excellent limitations on the application of the model, so we measured the training time consumption of our model SimAST-GCN and the other three baseline models. The results are shown in Figure 7.

In Figure 7, we can observe that SimAST-GCN consumes the least amount of time than other baseline models. We believe that there are two reasons for our excellent model training efficiency. First, our model has good parallelism capability. ASTNN requires node information to propagate from leaf nodes to root nodes, and this is a sequential process. The root node cannot be calculated when the information is not propagated, which leads to a long time-consuming model. DACE and TBRNN use a large number of RNN networks in their models, so they are not very good in terms of parallelism. The SimAST-GCN model we proposed uses only a small amount of RNN network in the model, and a large number of parallel GCN networks, which greatly improves the parallelism of the network, thereby greatly reducing the time consumption of model training and improving the training efficiency of the model. Second, we preprocess the data for all models. We move the extraction of AST, the simplifying of AST, and the extraction of node relationship graphs into the preprocessing process. Thus, the execution efficiency of the model is improved.

In conclusion, our model SimAST-GCN outperforms other models in terms of time efficiency.

\begin{table*}[t]
    \begin{center}
        \caption{Statistics of the tree size for the original and simplified versions of the AST.}
        \resizebox{\textwidth}{!}{ 
        \begin{tabular}{c|c|c c|c|c |c | c}
            \hline

            \hline
            Repository & Operation & Max token & Average token & Simplified rate & Average code token & Code token rate & Percentage increase\\
            \hline
            \multirow{2}{*}{accumulo} & Original & 3334 & 188.52 & \multirow{2}{*}{43.84\%} & \multirow{2}{*}{89.22} & 47.33\% & \multirow{2}{*}{36.95} \\
 & Simplified & 1949 & 105.87& & & 84.27\% & \\
\hline
\multirow{2}{*}{ambari} & Original & 2407 & 217.86 & \multirow{2}{*}{43.83\%} & \multirow{2}{*}{103.07} & 47.31\% & \multirow{2}{*}{36.92} \\
 & Simplified & 1339 & 122.36& & & 84.23\% & \\
\hline
\multirow{2}{*}{cloudstack} & Original & 5036 & 217.16 & \multirow{2}{*}{43.94\%} & \multirow{2}{*}{102.24} & 47.08\% & \multirow{2}{*}{36.91} \\
 & Simplified & 1790 & 121.73& & & 83.99\% & \\
\hline
\multirow{2}{*}{commons-lang} & Original & 1718 & 112.43 & \multirow{2}{*}{48.49\%} & \multirow{2}{*}{53.35} & 47.45\% & \multirow{2}{*}{44.67} \\
 & Simplified & 831 & 57.91& & & 92.12\% & \\
\hline
\multirow{2}{*}{flink} & Original & 1928 & 145.69 & \multirow{2}{*}{44.14\%} & \multirow{2}{*}{68.56} & 47.06\% & \multirow{2}{*}{37.19} \\
 & Simplified & 1069 & 81.38& & & 84.24\% & \\
\hline
\multirow{2}{*}{incubator-pinot} & Original & 3751 & 187.02 & \multirow{2}{*}{44.4\%} & \multirow{2}{*}{88.75} & 47.45\% & \multirow{2}{*}{37.89} \\
 & Simplified & 2388 & 103.99& & & 85.34\% & \\
\hline
\multirow{2}{*}{kafka} & Original & 2650 & 149.16 & \multirow{2}{*}{45.96\%} & \multirow{2}{*}{70.42} & 47.21\% & \multirow{2}{*}{40.15} \\
 & Simplified & 1265 & 80.6& & & 87.36\% & \\
\hline
\multirow{2}{*}{lucene-solr} & Original & 5467 & 242.65 & \multirow{2}{*}{43.94\%} & \multirow{2}{*}{113.96} & 46.96\% & \multirow{2}{*}{36.82} \\
 & Simplified & 3059 & 136.02& & & 83.78\% & \\
\hline
\multirow{2}{*}{shardingsphere} & Original & 434 & 73.14 & \multirow{2}{*}{43.11\%} & \multirow{2}{*}{34.86} & 47.66\% & \multirow{2}{*}{36.11} \\
 & Simplified & 247 & 41.61& & & 83.77\% & \\
\hline
            \multirow{2}{*}{Average} & Original & - & 170.41 & \multirow{2}{*}{44.5\%} & \multirow{2}{*}{80.49} & 47.28\% & \multirow{2}{*}{38.18} \\
             & Simplified & - & 94.61 & & & 85.46\% & \\
            \hline

            \hline
            
        \multicolumn{8}{l}{\small *Simplified rate $= 1 - $Simplified average token $/$ Original average token. } \\
        \multicolumn{8}{l}{\small *Code token rate = Average code token $/$ Average token (Original/Simplified).} \\
        \multicolumn{8}{l}{\small *Percentage increase = Simplified code token rate $-$ Original code token rate.} \\
        \end{tabular}
        }
    \end{center}
    
\end{table*}

\subsection{Discussion}
In this section, we will discuss the effectiveness of Simplifying AST. 
Table 11 lists the tree size before and after simplifying the AST. We can see that the average token of the tree drops from 170 to 94. We also notice that the proportion of code tokens in all tokens has also greatly increased (from 47\% to 85\%), the average percentage increase is about 38 percentage points. 
We believe that when the AST is not simplified, the proportion of extra nodes generated is too high, which will cause the model to focus too much on the information of the extra nodes, thereby ignoring the information of the code nodes themselves. Therefore, by simplifying the AST, on the one hand, we can strengthen the contact information between nodes, and on the other hand, we can increase the proportion of code nodes, so that the two kinds of node information can be better balanced. 
This is why simplified AST helps to predict the code review results. 

The reasons for simplifying AST are two-fold. First, after removing useless attribute nodes, the remaining attribute nodes represent the relationship between nodes, not the attributes of a specific node. For example, the ``modifiers" node in Figure 3(a) means that the ``static" node is a modifier node. However, the ``Method Declaration" node can unify its child nodes and shorten the distance between the code nodes, thereby strengthening the connection between the nodes. Second, a smaller number of nodes can reduce the computational cost and training overhead.

\section{Related Work}
\subsection{Source Code Representation}
In the field of software engineering, lots of code-related research needs to transform the code into a machine understandable form. Therefore, how to effectively represent the source code fragment is a significant challenge in the field of software engineering research.

Deep learning based methods have attracted much attention in learning representation of source code fragments. Raychev \cite{raychev2014code} uses n-gram and RNN model for the code completion task, and the main idea is to simplify the code completion problem to a natural language processing problem, that is, to predict the probability of a sentence. Allamanis \cite{allamanis2015suggesting} proposes a neural probabilistic language model for the method naming problem. They think that in similar contexts, the name tends to have similar embeddings.

However, with the deepening of the research, the researchers found that the structural information in the code is very important, so they began to study how to extract structural information in source code fragments. Mou \cite{mou2016convolutional} proposes a novel neural network (TBCNN), which is a tree-based model designed for programming language processing. They propose a convolution kernel used on the AST to capture the essential structural information. 
Lam \cite{lam2015combining} combines the project's bug-fixing history and the features built from rVSM and DNN for better accuracy in bug localization task. Huo \cite{huo2016learning} proposes a convolutional neural network NP-CNN, which is used to leverage both syntactic and semantic information. According to the bug report, NP-CNN learns unified features from the natural language and the source code fragments to predict potential buggy source code automatically. Wei \cite{wei2017supervised} proposes a methods called CDLH for functional clone detection, which is an end-to-end learning framework. CDLH exploits both syntactic and semantic information to learn hash codes for fast computation between different code fragments. Zhang \cite{8812062} proposes a method ASTNN. The main idea is to obtain better feature extraction ability by dividing the AST into sentence-level blocks.

Compared with these methods of serializing structural information into tokens for modeling, with the recent development of graph neural networks, many researchers have tried to directly use the original structural information for modeling instead of destroying the original structural information. Allamanis \cite{allamanis2017learning} represents source code fragments as graphs and uses different edge types to model semantic and syntactic relation information between different nodes. Zugner \cite{zugner2021language} combines the context and the structure information of source code and uses multiple programming languages as the dataset to improve results on every individual languages. 

Unlike the previous methods, our method SimAST-GCN not only strengthens the structural information, but also uses the graph convolution network to deeply integrate the structural information and semantic information to better represent the characteristics of the source code fragments.


\subsection{Automatic Code Review}
As an crucial part of software engineering, code review plays a pivotal role in the entire software life cycle. Code review determines the review results by convening other developers to understand, analyze and discuss the code. There is no doubt that this whole process requires many human resources. Therefore, many studies are devoted to reducing the consumption of human resources in code review process. Thongtanunam \cite{thongtanunam2015should} reveals that 4\%-30\% of reviews have code reviewer assignment problem. Thus, a code reviewer recommendation algorithm, File Path Similarity (FPS), was proposed to exploit the file location information to solve the problem. Zanjani \cite{zanjani2015automatically} proposed an approach called cHRev, which is used to recommend the best suitable reviewer to participate in a given review. The method makes the recommendation based on the contributions in their prior reviews. Xia \cite{xia2017hybrid} proposes a recommendation algorithm which leverages the implicit relations between the reviews and the historical reviews. So, they utilize a hybrid approach, combining the latent factor models and the neighborhood methods to capture implicit relations. They are all researching how to recommend suitable reviewers to improve the efficiency of code review.

Although there are so many code reviewer recommendation related works, it can only improve the efficiency of code review, but can not effectively reduce the human effort consumption of code review \cite{singh2017evaluating}. Rigby \cite{rigby2013convergent} finds that despite differences between items, many characteristics of the review process independently converge to similar values. They believe that this represents a general principle of code review practice. Therefore, we believe that since there are general principles of code review practice, then we can use existing deep learning techniques to learn these general principles. Just like the research of deep learning technology in other aspects of software engineering. Shi \cite{shi2019automatic} believes that automatic code review is a binary classification problem. Their model can learn the differences between the original file and the revised file to make the suggestion. Thus, they propose a novel model called DACE, which learns the revision features by exploiting a pairwise autoencoding and a context enrich module. 

However, the understanding of automatic code review is not the only one, Tufan \cite{tufan2021towards} proposes a method, learning the code changes recommended by reviewer, to implement them in the original code automatically. In other words, they are trying to make a map from the original code file to the revised code file, which is totally different from the opinion that Shi \cite{shi2019automatic} hold. In this paper, our understanding of automatic code review is the same as Shi \cite{shi2019automatic}, so we are more focused on optimizing the representation of the code and improving the accuracy of prediction.

\section{Conclusion}
In this paper, we first present AACR, a challenging dataset for automatic code review. Then, we propose a Simplified AST based Graph Convolutional Network (SimAST-GCN) to extract syntactic and semantic information from source code. SimAST-GCN first extract AST from the source code and simplifying the extracted AST. Then, SimAST-GCN uses Bi-GRU to enrich the semantic information and GCN to enrich the syntactic information. Finally, SimAST-GCN composes the representations from the original code file and the revised code file to predict the results. Experimental results on the AACR dataset showed that our proposed model SimAST-GCN outperforms state-of-the-art methods, including Token-based models and GCN-based models. Our code and experimental data are publicly available at \href{https://github.com/SimAST-GCN/SimAST-GCN}{https://github.com/SimAST-GCN/SimAST-GCN}.

\section*{Acknowledgment}
This work was partially supported by the National Natural Science Foundation of China (61772263, 61872177,
61972289, 61832009), the Collaborative Innovation
Center of Novel Software Technology and Industrialization,
and the Priority Academic Program Development of Jiangsu Higher Education Institutions.

\bibliographystyle{model1-num-names}

\bibliography{cas-refs}

\end{document}